\documentclass[english]{iopart}

\usepackage{graphicx}
\usepackage{url}
\usepackage{iopams} 

\expandafter\let\csname equation*\endcsname\relax
\expandafter\let\csname endequation*\endcsname\relax

\usepackage{amssymb}
\usepackage[latin1]{inputenc}
\usepackage{amsmath}
\usepackage{epsfig}

\newcounter{fig}
\begin{document}

{\Large Diagonal Ising susceptibility: elliptic integrals, 
modular forms and Calabi-Yau equations} 
\author{ M. Assis$^\dag$, S. Boukraa$||$, S. Hassani$^\S$, M. van Hoeij$^\ddag$, 
J-M. Maillard$^\pounds$  B.M. McCoy$^\dag$ }
\address{$^\dag$ CN Yang Institute for Theoretical Physics, State
  University of New York, Stony Brook, NY. 11794, USA} 
\address{$||$  \ LPTHIRM and D\'epartement d'A{\'e}ronautique,
 Universit\'e de Blida, Algeria}
\address{\S  Centre de Recherche Nucl\'eaire d'Alger, 
2 Bd. Frantz Fanon, BP 399, 16000 Alger, Algeria}
\address{\ddag Florida State University, Department of Mathematics,
1017 Academic Way, Tallahassee, FL 32306-4510 USA}
\address{$^\pounds$ LPTMC, UMR 7600 CNRS, 
Universit\'e de Paris, Tour 23,
 5\`eme \'etage, case 121, 
 4 Place Jussieu, 75252 Paris Cedex 05, France}

\begin{abstract}

We give the exact expressions of the partial
 susceptibilities $\, \chi^{(3)}_d$ and
$\, \chi^{(4)}_d$  for the diagonal susceptibility of the Ising model in
terms of modular forms and Calabi-Yau ODEs, 
and more specifically, 
 ${}_3F_2([1/3,2/3,3/2],\, [1,1];\, z)$ and
${}_4F_3([1/2,1/2,1/2,1/2],\, [1,1,1]; \, z)$ hypergeometric 
functions. By solving the connection problems
we analytically compute the behavior at all finite singular points for 
$\chi^{(3)}_d$ and $\chi^{(4)}_d$. We also give new results
for $\, \chi^{(5)}_d$. We see in particular, the emergence
of a remarkable order-six operator, which is such that
its symmetric square has a rational solution. 
These new exact results indicate that the 
linear differential operators occurring in the $\, n$-fold integrals
of the Ising model are not only ``Derived from Geometry'' (globally nilpotent),
but actually correspond to ``Special Geometry''
 (homomorphic to their formal adjoint).
This raises the question of seeing if these ``special geometry''
 Ising-operators, are ``special'' ones,
reducing, in fact systematically, to (selected, $\, k$-balanced, ...)
$\, _{q+1}F_q$ hypergeometric functions, or correspond to the more
general  solutions of Calabi-Yau equations.

\end{abstract}

\vskip .5cm


\noindent {\bf AMS Classification scheme numbers}: 34M55, 
47E05, 81Qxx, 32G34, 34Lxx, 34Mxx, 14Kxx 
\vskip .5cm


\section{Introduction}
\label{intro}

The magnetic susceptibility of the Ising model is defined in terms of
the two-point spin correlation function as
\begin{equation}
\label{bulk}
 k_BT \cdot \, \chi \, \, = \, \,  \, \, \sum_{M=-\infty}^{\infty}\sum_{N=-\infty}^{\infty}
\{\langle\sigma_{0,0}\sigma_{M,N}\rangle\, -{\mathcal M}^2\},
\end{equation}
where $\, {\mathcal M}$ is the spontaneous magnetization of the Ising model.

The exact analysis of the Ising model susceptibility is the most challenging
and important open question in the study of the Ising model today. This
study~\cite{bmw,wmtb} began in 1973-76  by means of summing 
the $n^{\mathrm{th}}$ particle
form factor contribution to the correlation function
$\langle\sigma_{0,0}\sigma_{M,N}\rangle$. In these papers, it was shown
that for $ \,T \,< \,T_c$ 
\begin{equation}
 k_BT \cdot \chi(t) \,  \,= \, \,  \,
(1-t)^{1/4} \cdot \, (1 +\sum_{n=1}^{\infty} \, {\tilde \chi}^{(2n)}(t)),
\end{equation}
where\footnote[1]{The classical interaction energy of the Ising model is
$-\sum_{j,k}\left(E^v\sigma_{j,k}\sigma_{j+1,k}
+E^h\sigma_{j,k}\sigma_{j.k+1}\right)$ 
where $j(k)$ specifies the row (column) of a square
lattice and the sum is over all sites of the lattice.}
  $\, t \,= \, \,(\sinh 2E^v/k_BT \sinh 2E^h/k_BT)^{-2}$
and for $T \,> \, T_c$ by
\begin{equation}
 k_BT \cdot \chi(t) \, \,  =  \, \, \, 
(1-t)^{1/4} \cdot \sum_{n=0}^{\infty} \, {\tilde \chi}^{(2n+1)}(t),
\end{equation}
where $\, t \,= \, \, (\sinh 2E^v/k_BT \sinh 2E^h/k_BT)^{2}$.

The $ \, {\tilde \chi}^{(n)}$ are given by $\, n$-fold integrals. 
In~\cite{wmtb}
the integrals for $\,{\tilde \chi}^{(1)}$ and
 $\,{\tilde \chi}^{(2)}$ were evaluated, and
since that time there have been many important
 studies~\cite{tracy,ongp,cgnp}, of the
behavior as $t\,\rightarrow \,1$, of 
the singularities in the
complex $t$-plane ~\cite{ongp,nickel1,nickel2} and 
the analytic properties of 
$\,{\tilde \chi}^{(n)}$ as a function of $t$ for the isotropic 
case~\cite{jm1}-\cite{2008-experimental-mathematics-chi}  for
$n= \, 3,\,4,\,5,\,6$. These studies are still ongoing.

More recently it was discovered~\cite{mccoy3} 
that if in (\ref{bulk}) the sum is
restricted to the spins on the diagonal
\begin{equation}
k_BT \cdot \chi_d(t)
 \,\,= \,\, \, \sum_{N=-\infty}^{\infty}\{\langle\sigma_{0,0}\sigma_{N,N}\rangle
-{\mathcal M}^2\}, 
\end{equation}
 the diagonal
susceptibility reads
\begin{equation}
 k_BT \cdot \chi_d(t)\, \,= \, \,\,  \,
(1-t)^{1/4} \cdot \left(1+\sum_{n=1}^{\infty}{\tilde \chi}_d^{(2n)}(t)\right),
\end{equation}
for $T\,<\,T_c$ and
\begin{equation}
 k_BT \cdot \chi_d(t)\,=\,\,\, 
(1-t)^{1/4} \cdot \sum_{n=0}^{\infty} \,{\tilde \chi}_d^{(2n+1)}(t),
\end{equation}
for $ \,T \,> \,T_c$. The $ \, {\tilde \chi}^{(n)}_d(t)$'s 
are $n$-fold integrals which have a
much simpler form than the integrals for $ \, {\tilde \chi}^{(n)}(t)$ but retain all
of the physically interesting properties of these integrals. 

For $T\,<\,T_c$, the integrals $ \, {\tilde \chi}^{(2n)}_d(t)$ read
\begin{eqnarray}
&&\hspace{-0.3in}{\tilde \chi}^{(2n)}_{d}(t)\,\, = \,\,\,  \, \,
 {{  t^{n^2}} \over {  (n!)^2 }} \, 
{{1 } \over {\pi^{2n} }} \cdot  
\int_0^1 \cdots  \,\int_0^1\prod_{k=1}^{2n}\,  dx_k  
\cdot   {1\, +t^n\, x_1\cdots x_{2n}\over
  1\,-t^n\, x_1 \cdots x_{2n}}
\nonumber \\ 
&&\hspace{-0.3in}\quad \quad \quad \quad \times
\prod_{j=1}^n\, \left({x_{2j-1}(1-x_{2j})(1-tx_{2j})\over 
x_{2j}(1-x_{2j-1})(1\, -t\, x_{2j-1})}\right)^{1/2}
\nonumber\\
&&\hspace{-0.3in}\quad \quad \quad \quad \times \prod_{1 \leq j \leq n}
\prod_{1 \leq k \leq n}(1\, -t\, x_{2j-1}\, x_{2k})^{-2}
 \nonumber\\
&&\hspace{-0.3in}\quad\quad \quad \quad  \times
\prod_{1 \leq j<k\leq n}(x_{2j-1}-x_{2k-1})^2\, (x_{2j}-x_{2k})^2, 
\label{chidm}
\end{eqnarray}
where $t$ is given by $ \, t \, = \,  \, (\sinh 2E^v/k_BT \sinh 2E^h/k_BT)^{-2}$.

For $T\,>\,T_c$, the integrals $ \, {\tilde \chi}^{(2n+1)}_d(t)$ read
\begin{eqnarray}
&&\hspace{-0.5in}{{\tilde \chi}}^{(2n+1)}_{d}(t) \,= \, \, {t^{n(n+1)}
 \over \pi^{2n+1} n!  (n+1)! } \cdot 
\int_0^1 \cdots \int_0^1 \, \, \prod_{k=1}^{2n+1}  dx_k
\nonumber\\
&& \times {1\, +t^{n+1/2}\,x_1\cdots x_{2n+1}\over
 1\, -t^{n+1/2}\, x_1\cdots x_{2n+1}} \cdot 
 \prod_{j=1}^{n}\, \Bigl((1-x_{2j})(1\,-t\,x_{2j})\cdot
 x_{2j}\Bigr)^{1/2}
 \nonumber \\
&&\times\prod_{j=1}^{n+1} \, 
\Bigl((1\,  -x_{2j-1})(1\,-t\, x_{2j-1}) \cdot x_{2j-1}\Bigr)^{-1/2} 
\nonumber\\
\label{chidp}
&&\times
\prod_{1\leq j\leq n+1}\prod_{\, \, \, 1\leq k \leq n}\, 
(1\, -t\, x_{2j-1}\, x_{2k})^{-2} 
\\
&&\times\prod_{1 \leq j<k\leq n+1}(x_{2j-1} -x_{2k-1})^2
\prod_{1\leq j<k\leq n}(x_{2j}-x_{2k})^2.
\nonumber
\end{eqnarray} 
where  $t\, = \, (\sinh 2E^v/k_BT \sinh 2E^h/k_BT)^2$.
For these $\, {{\tilde \chi}}^{(2n+1)}_{d}$ it will
 be convenient, in the following, to use the
variable   
$x\, \, = \, \, t^{1/2}\, = \, \sinh 2E^v/k_BT \sinh 2E^h/k_BT$.

In~\cite{mccoy3} we found that
\begin{equation}
{\tilde \chi}^{(1)}_d(x)\,\, = \, \,\,\,  \frac{1}{1\, -x}, \qquad
\hbox{and:}  \quad \quad \quad
{\tilde \chi}^{(2)}_d(t)\, =\,\, \,  {{1} \over {4}} \cdot \frac{t}{1 \, -t}, 
\end{equation}
and that $\, {\tilde \chi}_d^{(3)}(x)$ and $\, {\tilde \chi}_d^{(4)}(t)$ are 
solutions of differential equations of order 6 and 8.
The corresponding linear differential operators of each is a direct sum of
three factors.
In both cases, there was a differential equation which was not
solved in~\cite{mccoy3}. 

In this paper, we complete this study of  
$\, {\tilde \chi}_d^{(3)}(x)$ and 
$\, {\tilde \chi}_d^{(4)}(t)$ by solving all of the
differential equations involved. We then use the solutions of these
equations to analytically compute the singular behavior at all of the finite
singular points. In this way we are able to give analytic 
proofs of the results
conjectured in Appendix B of~\cite{mccoy3} by numerical means.

We split  the presentation of our results into two parts: the solution
of the differential equations and the use of the differential equations
to compute the behavior of $ {\tilde \chi}^{(3)}_d(x)$ 
and $ {\tilde \chi}^{(4)}_d(t)$ at
the singularities. The solution of the differential equations is
presented in section \ref{comput}
 for $ {\tilde \chi}^{(3)}_d(x)$ and in section \ref{computchi4gen} for
$ {\tilde \chi}^{(4)}_d(t)$. In particular we focus on the difficult
problem of solving a particular order-four operator,
to discover, finally, a  surprisingly simple result.
The linear differential equation for $\chi^{(5)}_d(x )$ 
is studied in section \ref{chi5}, yielding the emergence 
of a remarkable order-six operator.
The singular behaviors of  $\, {\tilde \chi}_d^{(3)}(x)$
 and $\, {\tilde \chi}_d^{(4)}(t)$ 
 are given in section \ref{singbehavchi3}
and \ref{singbehavchi4}, respectively. This 
analysis requires that the (global)
connection problem to be solved. The details of these computations
are given in appendices C and D. 
We conclude in section \ref{concl} with a
discussion of the emergence of $\, _{q+1}F_q$ hypergeometric functions, 
with all these previous results underlying 
modularity~\cite{SP4,LianYau} in the Ising model
through elliptic integrals, modular forms and
 Calabi-Yau ODEs~\cite{Almkvist,TablesCalabi}.

\section{Computations for ${\tilde \chi}^{(3)}_d(x)$}
\label{comput}

It was shown in~\cite{mccoy3} that $\, {\tilde \chi}^{(3)}_d(x)$
 is annihilated by an order-six
linear differential equation. The corresponding linear 
differential operator ${\cal L}^{(3)}_6$ is a
direct sum of irreducible linear differential
 operators (the indices are the orders):
\begin{eqnarray}
{\cal L}^{(3)}_6  \,\,= \, \, L^{(3)}_1 \oplus L^{(3)}_2 \oplus L^{(3)}_3.
\end{eqnarray}
The solution of ${\cal L}^{(3)}_6$ which is analytic at $\, x\, = \, \, 0$
is thus naturally decomposed as a sum:
\begin{eqnarray}
\label{chisum}
Sol( {\cal L}^{(3)}_6 ) \,\,= \,\,\, \, 
a_1^{(3)} \cdot {\tilde \chi}^{(3)}_{d;1}(x) \,
 +a_2^{(3)} \cdot {\tilde \chi}^{(3)}_{d;2}(x)\, 
+a_3^{(3)} \cdot {\tilde \chi}^{(3)}_{d;3}(x), 
\end{eqnarray}
where the  ${\tilde \chi}^{(3)}_{d;j}$  are analytic at $\, x \, = \, \, 0$.
The solutions ${\tilde \chi}^{(3)}_{d;1}(x)$ and ${\tilde \chi}^{(3)}_{d;2}(x)$    
were explicitly found in~\cite{mccoy3} to be
\begin{eqnarray}
\label{chi1}
&&\hspace{-0.85in} {\tilde \chi}^{(3)}_{d;1}(x)\,= \, \, \, 
\frac{1}{1-x}, \qquad \qquad \hbox{and:}  \\
\label{chi2}
&&\hspace{-0.85in} {\tilde \chi}^{(3)}_{d;2}(x) 
= \,  \frac{1}{(1-x)^2} \cdot \, _2F_1([1/2,-1/2],\, [1]; \, x^2) 
  -\frac{1}{1-x}\cdot \, _2F_1([1/2,1/2],\,[1]; \, x^2),  
\end{eqnarray}
where one notes the occurrence of
 ${\tilde \chi}^{(1)}_d(x) =\, {\tilde \chi}^{(3)}_{d;1}(x)$
 in $\, {\tilde \chi}^{(3)}_{d}(x)$.
The last term, ${\tilde \chi}^{(3)}_{d;3}(x)$, is annihilated by the order-three
linear differential operator
\begin{eqnarray}
\label{l33}
&& L_3^{(3)}\, = \,\,\,\,\,\, D_x^{3}\, \, \, 
+  {{3} \over {2}}\,{\frac { n_2(x)}{ d(x)}} \cdot D_x^{2} \,\,
\,  +{\frac {n_1(x) }{  (x+1) (x-1)  \cdot  x \cdot d(x) }} \cdot  D_x 
\nonumber \\
&&\qquad \quad \quad  
+{\frac { n_0(x)}{ (x+1)  \, (x-1)^2  \cdot  x \cdot d(x)}}, 
\nonumber
\end{eqnarray}
where:
\begin{eqnarray}
&&\hspace{-0.2in}d(x) \, = \, \,\,
(x+2)  \, (1+2\,x)  \, (x+1)  \, (x-1)  \, (1+x+x^2)\cdot  x, 
 \nonumber \\
&&\hspace{-0.2in}n_0(x) \, = \, \,\,
2\,{x}^{8} +8\,{x}^{7} -7\,{x}^{6} -13\,{x}^{5}
-58\,{x}^{4} -88\,{x}^{3} -52\,{x}^{2}-13\,x+5,   
\nonumber \\
&&\hspace{-0.2in}n_1(x) \, = \, \,\, 
14\,{x}^{8} +71\,{x}^{7} +146\,{x}^{6} +170\,{x}^{5} +38\,{x}^{4}  
\nonumber \\
&&\hspace{-0.2in}\quad \quad \quad \quad \quad \quad 
-112\,{x}^{3} -94\,{x}^{2} -19\,x+2.   
\nonumber \\
&&\hspace{-0.2in}n_2(x) \, = \, \,\, 8\,{x}^{6} +36\,{x}^{5} +63\,{x}^{4}
+62\,{x}^{3} +21\,{x}^{2} -6\,x -4. 
\end{eqnarray}

\vskip 0.2cm

\vskip 0.3cm

The linear differential operator $\, L_3^{(3)}$
has the following regular singular points and exponents
($z$ denotes the local variable $x-x_s$ of the expansion
 around a singular point $\, x_s$):
\begin{eqnarray}
\label{l33exp}
1+x+x^2 = 0,&\quad\quad \rho = \, 0,\, 1,\, 7/2 & \rightarrow 
 \quad \quad \quad \,\,z^{7/2},  
\nonumber \\
x = ~~ 0&\quad \quad \rho \, =  \,0, \, 0, \, 0 & \rightarrow 
\quad \quad\quad \,\, \ln(z)^2\,\,\,\, terms, 
\nonumber \\
x = ~~1 &\quad \quad  \rho  \,= \, -2, \, -1, \, 1 & \rightarrow 
\quad \quad \quad \,\, z^{-2}, \,\, z^{-1}, \\ 
x = -1&\quad \quad \rho = 0, 0, 0 & \rightarrow 
\quad \quad \quad \,\, \ln(z)^2\,\,\,\, terms,  
\nonumber \\
x \, = \, ~\infty& \quad \quad \rho  \,= \, 1, \, 1, \, 1 & \rightarrow 
\quad \quad \quad \,\, \ln(z)^2\,\, \,\,terms.  
\nonumber
\end{eqnarray}
The last column shows the maximum $\ln(z)$-degree
 occurring in the formal solutions of
$ \, L_3^{(3)}$, $z$ being the local variable of the expansion.
 
The singularities at $ \,x \, =\, -2,\, -1/2$ are apparent.

By use of the command dsolve in Maple, we found in~\cite{jm6} that
the solution to
 $\, L^{(3)}_3[{\tilde \chi}^{(3)}_{d;3}] \,=\,\, 0$ which is analytic at $x=\, 0$ is 
\begin{eqnarray}
\label{chi31}
\hspace{-0.2in} {\tilde \chi}^{(3)}_{d;3}(x)\,\,
=\, \,\, \frac{(1\, +2x)\cdot(x\, +2)}{(1-x) \cdot (x^2+x+1)} 
 \, \cdot \,     {}_3F_2([1/3,2/3,3/2],[1,1]; \, Q),
\end{eqnarray}
where the pullback $\, Q$ reads:
\begin{equation}
\label{qdef}
Q\,\, =\,\,\,\,\,
\frac{27}{4}\,\frac{(1+x)^2 \cdot x^2}{(x^2+x+1)^3}. 
\end{equation}

Now the coefficients $a_i^{(3)}$ in the  
sum decomposition (\ref{chisum}) of ${\tilde \chi}^{(3)}_{d}(x)$,
 can be fixed by expanding and matching the rhs of (\ref{chisum}) 
with the expansion of ${\tilde \chi}^{(3)}_{d}(x)$. This gives
\begin{eqnarray}
\label{chisum2text}
{\tilde \chi}^{(3)}_d(x)\, \,=\,\,\, \,\,
\frac{1}{3} \cdot {\tilde \chi}^{(3)}_{d;1}(x)
\,\, +\frac{1}{2} \cdot {\tilde \chi}^{(3)}_{d;2}(x)
\,\,-\frac{1}{6} \cdot {\tilde \chi}^{(3)}_{d;3}(x).
\end{eqnarray}

By use of a family of identities on $\, _3F_2$
 hypergeometric functions~\cite{Prudnikov}
(see eqn. 27 page 499) the expression (\ref{chi31}) 
of  ${\tilde \chi}^{(3)}_{d;3}(x)$
 reduces to
\begin{eqnarray}
\label{Prud}
&&\hspace{-0.5in}  {\tilde \chi}^{(3)}_{d;3}(x)  
\, \, = \,  \,\, 
\frac{(1 \, +2x)\cdot (x\, +2)}{(1-x)\cdot (x^2+x+1)} \, \cdot 
\left[ \,_2F_1([1/6,1/3],\,[1]; \, Q)^2\right.
\\
&&\hspace{-0.2in} \quad \quad  \quad 
 \left.+\, \frac{2Q}{9}\, \cdot \,  
 _2F_1([1/6,1/3],\,[1];\,Q) \,\cdot \,  
 _2F_1([7/6,4/3],\,[2];\, Q) \right].
\nonumber
\end{eqnarray}

It is instructive, however, to discuss further the reason why
${\tilde \chi}^{(3)}_{d;3}(x)$  has this solution in terms 
of ${}_2F_1$ functions.

\subsection{Differential algebra structures and modular forms}
\label{compu}

From a differential algebra viewpoint, the
linear differential operator $\, L_3^{(3)}$
can be seen to be homomorphic\footnote[3]{
For the notion of differential operator equivalence see~\cite{vdP}
and~\cite{Homomorphisms}.}
 to its formal adjoint:
\begin{eqnarray}
\label{homoadj}
L_3^{(3)} \cdot   adjoint(T_2 ) 
\, \,\, \, \, = \, \, \, \,\, \,T_2  \cdot adjoint(L_3^{(3)}), 
\end{eqnarray}
where:
\begin{eqnarray}
&&\hspace{-0.5in}T_2 \, \,  \, \, = \,  \,\, \, \,  \,
 {{(1+x+x^2) } \over { (1\, -x)^4}} \cdot D_x^2  \,
 \, \,  \,+ \, {{ m_1(x)} \over {
 (x\, +1) \, (x \, -1)^5 \, (2 \, x+1) \, (x+2) \cdot x }} \cdot D_x 
\nonumber \\
&&\hspace{-0.1in} \quad 
-\, {{1} \over {4}} \cdot {{ m_0(x)} \over {
(2 \, x+1) \, (x+2) \, (x \, +1) \, (1+x+x^2) \, (x \, -1)^6 \cdot x
 }}, 
\end{eqnarray}
and where:
\begin{eqnarray}
&& m_1(x)  \,  \, = \,  \, \, \,2 \, x^6\, -6 \, x^5\, 
-53 \, x^4\, -92 \, x^3\, -81 \, x^2\, -34 \, x\, -6,
 \nonumber \\
&&  m_0(x)  \,  \, = \, \,  \, \, 8 \, x^8 \, -4 \, x^7-222 \, x^6 \,
 -769 \, x^5\, -1153 \, x^4\,
\nonumber \\
&& \qquad  \quad \quad  -1341 \, x^3\, -1129 \, x^2\, -490 \, x\, -84. 
 \nonumber
\end{eqnarray}
Related to (\ref{homoadj}) is the property that 
the symmetric square\footnote[2]{In general, for an irreducible
 operator homomorphic to its adjoint, a rational solution occurs
 for the symmetric square (resp. exterior  square)
of that operator when it is of odd (resp. even) order.}
 of $\, L_3^{(3)}$ actually has a (very simple) 
rational solution $\, R(x)$. It thus factorises into an (involved) order-five
linear differential operator and an order-one  operator
 having the rational solution:
\begin{eqnarray}
\label{exteri}
\hspace{-0.5in} R(x) \,  \, =  \, \,  \, {{ 1\, +x \, +x^2} \over {(x-1)^4 }}, 
\quad \quad   
Sym^2( L_3^{(3)}) \, \, = \, \, L_5 \cdot 
\Bigl(D_x \, - \, {{d} \over {dx}} \, \ln(R(x))  \Bigr).  
\end{eqnarray}
In a forthcoming publication
we will show that the homomorphism of an operator with its adjoint
naturally leads to a rational solution for its {\em symmetric square  
or exterior square} (according to the order of the operator). 

Relation (\ref{homoadj}), or the fact that its symmetric square 
has a rational solution,  means that this operator is not only
 a globally nilpotent operator~\cite{jm6}, but it
 corresponds to ``Special Geometry''.
In particular it has a ``special'' differential Galois group~\cite{Katz}.
We will come back to this crucial point below, in section \ref{computchi4}
(see (\ref{intertwi})). 

The operator $\, L_3^{(3)}$ is in fact homomorphic
 to the symmetric square
of a second order linear differential operator\footnote[3]{Finding $X_2$ (or   
an operator equivalent to it) can be done by downloading
the implementation~\cite{ISSAC}.}
\begin{eqnarray}
\label{X2}
\hspace{-0.5in} X_2 \,\, = \, \,\, \, D_x^2 \,\, 
+{{1} \over {2}} \cdot {{(2\,x\, +1) \cdot (x^2+x+2)
} \over {(1+x+x^2)  \cdot (1+x) \cdot x}}  \cdot D_x 
\, \, - {{3} \over {2}} \cdot  {{1} \over {(1+x+x^2)^2}},
\end{eqnarray}
since one has the following simple operator equivalence~\cite{vdP}
with two order-one intertwinners:
\begin{eqnarray}
\label{M1}
&&\hspace{-0.6in}L_3^{(3)} \cdot   M_1 
 \, \,\,  \, = \,\,  \, \, \,  N_1  \cdot Sym^2(X_2), 
\qquad \qquad \quad \hbox{with:} 
 \\
&&\hspace{-0.6in}M_1   \, \, = \, \, \,\, 
  {{(1+x) \cdot x } \over { (1 -x)^2}} \cdot D_x \,\,\,
 + \, \,{{1} \over {2}} \cdot 
 {{(1\, +2\, x) \cdot (x+2) } \over {(1+x+x^2) \cdot (1 -x)}}, 
\nonumber \\
&&\hspace{-0.6in}N_1   \, \, = \, \, \,\,  
 {{(1+x) \cdot x } \over {(1 -x)^2 }} \cdot D_x 
 \,\, - \, \, 
{{1} \over {2}} \cdot
{{24 \, x^5+15 \, x^4+8 \, x^6-10 \, x^3-69 \, x^2-60 \, x-16 
} \over {(1 -x)^3 (1\, +2\, x) \, (x\, +2) \,(1+x+x^2)}}.
\nonumber
\end{eqnarray}
The second order operator $\, X_2$ is not homomorphic
 to the second order operators 
associated with the  complete elliptic integrals of the first or second kind.
However, from  (\ref{homoadj}) and (\ref{exteri}), we 
expect  $\, X_2$ to be ``special''.
This is confirmed by the fact that the
 solution $\, Sol(X_2, \, x)$ of $\, X_2$, analytical
at $\, x \, = \, \, 0$ has the integrality 
property\footnote[1]{See also the concept of 
``Globally bounded'' solutions of linear differential equations by
 G. Christol~\cite{Christol}.}:
 if one performs  a simple rescaling 
$\, x \, \rightarrow \, 4 \, x$ the series expansion 
of this solution has {\em integer} coefficients:
\begin{eqnarray}
&&\hspace{-0.2in} Sol(X_2, \,4\, x)\,\, = \, \,\, \,\,\,
1 \, +6\, x^2\, -24 \, x^3\, +60 \, x^4\,  -96 \, x^5\,
\nonumber \\
&& \qquad  \, +120\, x^6\, -672\, x^7\,
 +5238\, x^8\, -25440\, x^9\, +\,\, \cdots 
\end{eqnarray}

From this {\em integrality} property~\cite{jm9,Kratten}, we
 thus expect the solution of $\, X_2$ 
to be associated with a modular form, and thus, we expect this solution 
to be a $\, _2F_1$ up to not just one, {\em but two pullbacks}. 
Finding these pullbacks
is a difficult task, except if the  pullbacks are rational functions.
Fortunately we are in this simpler case of rational 
pullbacks, and consequently, 
we have been able to find the solution~\cite{jm6} to
 deduce that the third order operator $\, L^{(3)}_3$
is  $_3F_2$-solvable or $_2F_1$-solvable,
up to a Hauptmodul~\cite{Harnad} pullback 
(see (\ref{chi31}), (\ref{Prud})).

We can make the modular form character of  (\ref{chi31}), (\ref{Prud}),
which is already quite clear from the Hauptmodul form
of (\ref{qdef}), very explicit
by introducing another rational expression, similar to (\ref{qdef}):
\begin{eqnarray}
\label{qdef2}
  Q_1(x) \,\,   = \, \,\, \,  
{{27\, x\cdot (1+x) } \over { (1\, +\, 2\, x)^6}}.
\end{eqnarray}

The elimination of $\, x$ between $\, Q \, = \, \, Q(x)$ and
$\, Q_1 \, = \, \, Q_1(x)$
(see (\ref{qdef}), (\ref{qdef2})) gives 
a polynomial relation (with integer coefficients)
$\, \Gamma(Q, \, Q_1)\, = \, \, 0$, where 
the algebraic curve $\, \Gamma(u, \, v)\, = \, \, 0$ 
which is, of course, a rational curve, is, in fact a 
{\em modular curve} already encountered~\cite{jm9} with an order-three 
operator $\, F_3$ which emerged in (the non-diagonal)
$\, {\tilde \chi}^{(5)}$ (see~\cite{jm7}): 
\begin{eqnarray}
\label{modularcurv}
&&-4 \, u^3 \, v^3 \, +12 \, u^2 \, v^2 \cdot  (v+u) \, 
-3 \, u \, v \cdot  (4 \, v^2+4 \, u^2\, -127 \, u \, v) \, 
\nonumber \\
&&\qquad \quad  +4 \, (v+u) \cdot  (u^2+v^2 \,
 +83 \, u \, v) \, -432 \, u \, v
\,\,\,\,  = \,\,\, \, \, 0.
\end{eqnarray}

The hypergeometric functions we encounter in (\ref{Prud}),  
in the expression of the solution of $\, L^{(3)}_3$, have 
actually {\em two possible pullbacks}
as a consequence of the remarkable identity on 
the {\em same} hypergeometric function\footnote[2]{Along this line
 see for instance~\cite{Vidunas2011}.}:
\begin{eqnarray}
\label{bingo}
&& (1\, +2 \, x)
\cdot 
\, _2F_1\Bigl([{{1} \over {6}}, \,{{1} \over {3}}], \, [1]; \,Q(x)\Bigr)
\nonumber \\
&& \qquad \, \, = \, \, \, \,\, (1+x+x^2)^{1/2}
\cdot 
 \, _2F_1\Bigl([{{1} \over {6}}, \,{{1} \over {3}}], \, [1]; \,Q_1(x)\Bigr). 
\end{eqnarray}
Other rational parametrizations and pullbacks 
can also be introduced, as can be seen in \ref{miscell}.
Relation (\ref{bingo}) on $\, _2F_1$ yields other remarkable relations
on the  $\, _3F_2$ with the two pullbacks $\, Q$ (see (\ref{chi31}))
and $\, Q_1$: their corresponding 
order-three linear differential operators are homomorphic. 
Consequently one deduces,
for instance, that  $\, {}_3F_2([1/3,2/3,3/2],[1,1]; \, Q_1)$ 
is equal to the action 
of the second order operator $\, U_2$ on $\, {}_3F_2([1/3,2/3,3/2],[1,1]; \, Q)$:
\begin{eqnarray}
\label{U2}
&&\hspace{-0.8in}(x^2\, +x \, +1)^3 \cdot (1\, -8\, x\, -8\, x^2) \cdot \,
{}_3F_2([1/3,2/3,3/2],[1,1]; \, Q_1)  
\nonumber \\
&&\hspace{-0.8in}\quad \quad \quad 
\, \, = \, \, \, \, - \, U_2\Bigl[ {}_3F_2([1/3,2/3,3/2],[1,1]; \, Q) \Bigr],
\qquad \qquad \quad \hbox{where:} 
  \\
&&\hspace{-0.5in}U_2 \,  \, = \, \, \,
 p(x) \cdot \Bigl(D_x^2 \,  \, -\, 2 \cdot {{d} \over {dx}} \ln\Bigl( 
{{x^2\, +x \, +1 } \over {(x\, -1)\,(x\, +2)\, (1+2\, x) }} \Bigr) \cdot D_x  \Bigr)
 \nonumber \\
&&\hspace{-0.5in}\qquad \qquad  \, + \, \, 
\Bigl({{1+2\, x } \over {x^2\, +x \, +1 }}   \Bigr)^2 \cdot q(x),
\quad  \qquad \qquad \hbox{with:} 
\nonumber \\
&&\hspace{-0.5in}p(x) \,  \, = \,  \, \, x^2 \cdot \,  (1+x)^2\, (1+2\, x)^2 \,
 (1+8\, x+12\, x^2+8\, x^3+4\, x^4), 
\nonumber \\
&&\hspace{-0.5in}
q(x)  \, \,  = \, \, \, 8 \, x^{10}\, +40 \, x^9\, +81 \, x^8\,
 +84 \, x^7\, +24 \, x^6\, -54 \, x^5\,-63 \, x^4 
\nonumber \\
&&\hspace{-0.5in} \qquad \quad \, -18 \, x^3\, -2 \, x\, -1.
\end{eqnarray}
thus generalizing the simple automorphic relation (\ref{bingo}).

The {\em  modularity} of these functions can also be seen from the fact that 
the series expansion of (\ref{chi31}), (\ref{Prud}), or (\ref{bingo}) 
have {the \em integrality property}~\cite{jm9}.
If one performs a simple rescaling
 $\, \, x \, \rightarrow  \, \, 4 \, x$, 
their series expansions actually
 have {\em integer coefficients}~\cite{jm9,Kratten}:
\begin{eqnarray}
&&\hspace{-0.6in}{\tilde \chi}^{(3)}_{d;3}(4 \, x)\, \,\,  = \, \, \,\,
2 \, +20 \, x \, +104 \, x^2 \,+560 \, x^3 \,+2648 \, x^4 \,+12848 \, x^5
\nonumber \\
&&\hspace{-0.2in} \quad  \,+58112 \, x^6 \,
+267776 \, x^7 \,+1181432 \, x^8 \,+5281328 \, x^9 \, +  \,\, \cdots,  
\end{eqnarray}
or
\begin{eqnarray}
&&\hspace{-0.6in} _2F_1\Bigl([{{1} \over {6}}, \,{{1} \over {3}}], \, [1]; 
\,Q(x)\Bigr)[x \, \rightarrow \, 4 \cdot x]
\, \,\,\,  = \, \,\,\, 1\, +6\, x^2\, -24\, x^3\, +60\, x^4\, -96\, x^5\, 
\nonumber \\
&&\hspace{-0.3in} \quad \quad 
 +120\, x^6\,\, -672\, x^7 +5238\, x^8\, -25440\, x^9\, +81972\, x^{10}
 +  \,\, \cdots, 
\end{eqnarray}
which can be turned into {\em positive integers}
 if we also change $\, x$ into $\, -\, x$.

This provides more examples
of the almost quite systematic occurrence~\cite{jm9} in the Ising model,
of (globally nilpotent~\cite{Baldassarri}) linear differential operators 
associated with {\em elliptic curves}, either because 
one gets straightforwardly elliptic integrals,
or  because one gets  operators 
associated with {\em modular forms}. For the diagonal 
susceptibility of the Ising model, are we also going to see
the emergence of {\em Calabi-Yau}-like 
operators~\cite{Almkvist,TablesCalabi}
as already discovered in  (the non-diagonal)
$\, {\tilde \chi}^{(6)}$ (see~\cite{jm9}) ? 

\section{Computations for ${\tilde \chi}^{(4)}_d(t)$}
\label{computchi4gen}

We now turn to the computation of $\, {\tilde \chi}^{(4)}_d(t)$, 
whose differential operator 
${\cal L}^{(4)}_8$ is of order eight and is a direct sum of three
irreducible differential operators~\cite{mccoy3}:
\begin{eqnarray}
{\cal L}^{(4)}_8  \,\,= \, \, \,\, \, 
L^{(4)}_1  \, \oplus \, L^{(4)}_3 \, \oplus \, L^{(4)}_4.
\end{eqnarray}
The solution of ${\cal L}^{(4)}_8$ analytic at $\, t \, = \, \, 0$, is 
thus naturally decomposed as a sum:
\begin{eqnarray}
\label{chisumL8}
Sol( {\cal L}^{(4)}_8  ) \,\,= \, \,\, \, \, 
a_1^{(4)} \cdot {\tilde \chi}^{(4)}_{d;1}(t) \,
 +a_2^{(4)} \cdot {\tilde \chi}^{(4)}_{d;2}(t)\,
+a_3^{(4)} \cdot {\tilde \chi}^{(4)}_{d;3}(t). 
\end{eqnarray}

The solutions ${\tilde \chi}^{(4)}_{d;1}(t)$ 
and ${\tilde \chi}^{(4)}_{d;2}(t)$    
were explicitly found\footnote[1]{The first line in (\ref{4chi2})
 can, for instance, be found by directly using the command
dsolve in Maple, and the second line follows by use of identity 520 on
page 526 of~\cite{Prudnikov}. This result is also easily obtained by
using Maple to directly compute the homomorphisms between an 
order-three operator
and the operator which annihilates ${}_2F_1([1/2,1/2],[1];t)$. 
} to be~\cite{mccoy3}
\begin{eqnarray}
&&\label{4chi1}
{\tilde \chi}^{(4)}_{d;1}(t)\, =\, \,\, \frac{t}{1-t}, 
\qquad \qquad \quad \hbox{ and:} \\
\label{4chi2}
&&{\tilde \chi}^{(4)}_{d;2}(t)
\,\, =\,\,\,\,\,\,\, {{9} \over {8}} \cdot 
{{(1\,+t) \cdot t^2} \over {(1\, -t)^5}}
 \cdot \,  _3F_2\Bigl([{{3} \over {2}}, \,
 {{5} \over {2}}, \, {{5} \over {2}}], \,[3, \, 3];
 \, {{ -4 \, t} \over {(1\, -t)^2}}   \Bigr)
\nonumber \\   
&&=\,\,\, \frac{1+t}{(1-t)^2} \cdot \, _2F_1([1/2,-1/2],[1];\, t)^2
\, \, -\, _2F_1([1/2,\,1/2],[1];\,t)^2 
\nonumber\\
&&\quad \quad -\frac{2t}{1-t} \cdot\, _2F_1([1/2,1/2],[1];\,t )\, 
 \cdot\, _2F_1([1/2,-1/2],[1];\,t).
\end{eqnarray}
Here, again, one notes the occurrence of ${\tilde \chi}^{(4)}_{d;1}(t)$
 which is ${\tilde \chi}^{(2)}_{d}(t)$
up to a normalization factor.
One should be careful that the $\, _3F_2$
 closed form (\ref{4chi2}) for $\, {\tilde \chi}^{(4)}_{d;2}(t)$,
 together with the previous exact result (\ref{chi31}),
may yield a $\, _{q+1}F_q$ with a rational pullback prejudice
which has no justification for the moment. 

 Similar to  $\, L_3^{(3)}$, 
the order-three operator for $\, {\tilde \chi}^{(4)}_{d;2}(t)$,
 is homomorphic to its adjoint and
its symmetric square has a simple rational function solution. The exact 
expressions (\ref{4chi2}) for $\, {\tilde \chi}^{(4)}_{d;2}(t)$
 are obtained in a similar way 
to the solution (\ref{chi31}), (\ref{Prud})
of $\, L_3^{(3)}$ in the previous section. We first find~\cite{ISSAC} 
 that the corresponding
linear differential  operator is homomorphic to the symmetric square of
a second order operator, which turns out 
to have complete elliptic integral solutions.
The emergence in (\ref{4chi2}) of a $\, _3F_2$ hypergeometric function 
 with the selected\footnote[2]{The fundamental role played by such specific 
pullbacks as  {\em isogenies of elliptic curves}
 has been underlined in~\cite{Renorm}.}
rational pullback $\, -4\, t/(1-t)^2$ is totally reminiscent (even if it is not
exactly of the same form) of 
Kummers's quadratic relation~\cite{Vidunas2011,Renorm},  
and its generalization to $\, _3F_2$ hypergeometric functions 
(see the relations (4.12),  (4.13) in~\cite{Bailey},
 and (7.1) and (7.4) in~\cite{Whipple}), for example:
\begin{eqnarray}
\label{Whipp}
&&\hspace{-0.8in}_3F_2\Bigl([1\, +\, \alpha \, -\beta \, - \, \gamma, 
\, {{\alpha} \over {2}}, \,{{\alpha\, +1} \over {2}}],\, 
[1\, +\, \alpha \, -\beta, \,\, 1\, +\, \alpha \, - \, \gamma];\,
 {{ -4 \, t} \over {(1\, -t)^2}} \Bigr)
\nonumber \\
&&\hspace{-0.5in}\quad \,\, = \,\,\,\,\,\, (1\, -\, t)^{\alpha} \cdot \, \,  
_3F_2([\alpha, \, \beta, \, \gamma], \, 
[1\, +\, \alpha \, -\beta, \,\, 1\, +\, \alpha \, - \, \gamma];\,t).
\end{eqnarray}
which relates {\em different}\footnote[8]{Note that the 
Saalsch{\" u}tzian difference (\ref{diff}) 
(see below) of the $\, _3F_2$ at the lhs of (\ref{Whipp})
 is  independent of $\, \alpha$, $\, \beta$, $\, \gamma$  and is
equal to
 $\, 1/2$, in contrast with the rhs.}
  $\, _3F_2$ hypergeometric functions.
In fact, and similar to (\ref{U2}), we do have 
an equality between the $\, _3F_2$ hypergeometric
function with the pullback $\, u\, = \, \,  \, -4\, t/(1 \, -t)^2$ 
and the {\em same} $\, _3F_2$ hypergeometric,
 where the pullback has been changed\footnote[5]{This amounts to changing $\, t$
into $\, 1\, -t$.} into $\, v\, = \, \, -4\, (1 \, -t)/t^2$,  
\begin{eqnarray}
\label{37}
\hspace{-0.8in}_3F_2\Bigl([{{3} \over {2}}, \, 
{{5} \over {2}}, \, {{5} \over {2}}], \,[3, \, 3];
 \, {{ -4 \, (1-t)} \over {t^2}}   \Bigr) \, = \, \, \, 
  V_2\Bigl[\, _3F_2\Bigl([{{3} \over {2}}, \, 
{{5} \over {2}}, \, {{5} \over {2}}], \,[3, \, 3];
    \, {{ -4 \, t} \over {(1\, -t)^2}}   \Bigr)\Bigr],
\end{eqnarray}
where $\, V_2$ is a second-order operator similar to the
 one in (\ref{U2}). The elimination
 of $\, t$ in these two pullbacks, gives the simple genus zero curve
\begin{eqnarray}
u^2 \, v^2 \,\,\,\, -48 \, v \, u \,\,\, +64 \cdot \, (u + \, v) 
\,\, \,\,\, = \,\,\, \,\, \, 0,  
\end{eqnarray}
reminiscent of the simplest 
{\em modular equations}~\cite{modularequations}. This genus zero curve can also be
simply parametrized with $\, u\, = \, \,  \, -4\, t/(1 \, -t)^2$
 and\footnote[9]{This amounts to 
changing $\, t$ into $\, -t/(1-t)$ or $\, -1/(1-t)$.} 
$\, v \, = \, \,  \,4\, t \cdot (1\, -t)$. Again, one gets an identity, similar
to (\ref{37}), with another order-two intertwinner $\, {\cal V}_2$:
\begin{eqnarray}
\label{39}
\hspace{-0.9in}_3F_2\Bigl([{{3} \over {2}}, \,
 {{5} \over {2}}, \, {{5} \over {2}}], \,[3, \, 3];
 \, 4 \, t \cdot (1-t)   \Bigr) \, = \, \, \, 
  {\cal V}_2\Bigl[\, _3F_2\Bigl([{{3} \over {2}}, 
\, {{5} \over {2}}, \, {{5} \over {2}}], \,[3, \, 3];
    \, {{ -4 \, t} \over {(1\, -t)^2}}   \Bigr)\Bigr].
\end{eqnarray}

\subsection{Computation of ${\tilde \chi}^{(4)}_{d,3}(t)$}
\label{computchi4}

The third term $\, {\tilde \chi}^{(4)}_{d;3}(t)$ in the sum (\ref{chi4exp}) 
is the solution analytic at $\, x= \, 0$ of the order-four
 linear differential operator
\begin{eqnarray}
\label{l44}
&&\hspace{-0.5in}L_4^{(4)}\,\,  =\, \,\,\, \,  D_t^{4}\,\,\, +{\frac { n_3(t) }{ 
(t+1)  \cdot d_4(t)
 }}  \cdot D_t^{3}
 \,  \,  + 2\,{\frac { n_2(t)}{
 (t^2-1) \cdot t \cdot d_4(t)
}} \cdot  D_t^{2}
 \nonumber \\
&&\qquad  \, + 2\,{\frac { n_1(t)}
{  (t^2-1) \cdot t \cdot d_4(t)  }} \cdot  D_t 
 \,\,\,  -3\,{\frac { \left( t+1 \right)^{2}}
{  (t-1) \cdot t^2 \cdot d_4(t)}},
\end{eqnarray}
where:
\begin{eqnarray}
&&\hspace{-0.8in}d_4(t) \,\, = \, \, \, \, ({t}^{2}-10\,t +1) \cdot \, (t-1) \cdot t, 
\quad \quad \quad n_1(t) \,\, = \, \, \, \,  {t}^{4}
-13\,{t}^{3}-129\,{t}^{2}+49\,t\,-4, 
 \nonumber \\
&&\hspace{-0.8in}n_2(t) \,\, = \, \, \, \,  5\,{t}^{5}
-55\,{t}^{4}-169\,{t}^{3}+149\,{t}^{2}-28\,t\,+2, 
 \nonumber \\
&&\hspace{-0.8in}n_3(t) \,\, = \, \, \, \,  7\,{t}^{4}
-68\,{t}^{3}-114\,{t}^{2}+52\,t\,-5. 
\nonumber
\end{eqnarray}
The operator $\, L_4^{(4)}$ has the following regular singular points and
exponents 
\begin{eqnarray}
\label{l44ex}
t = ~~0,& \quad \quad \quad \rho = 0, 0, 0, 1 & \quad \quad   \rightarrow \quad 
\quad \quad  \ln(z)^3\,\, \,\, \, \,  terms, \nonumber \\
t = ~~1,& \quad\quad \quad  \rho = -2, -1, 0, 1 &
\quad \quad \rightarrow \quad \quad \quad 
 z^{-2},\, \, z^{-1},\, \, \ln(z)\,\,\,\, \,  \, term,
 \nonumber \\
t = -1,& \quad \quad\quad  \rho = 0, 1, 2, 7 & \quad \quad \rightarrow 
\quad \quad \quad   z^7\, \ln(z)\,\,\, \,  \, term, \\
t = ~\infty, & \quad \quad \quad \rho = 0, 0, 0, 1 &\quad \quad \rightarrow 
\quad \quad  \quad  \ln(z)^3 \,\,\,\, \,  \, terms. \nonumber
\end{eqnarray}
The last column shows the maximum $\ln(z)$-degree
 occurring in the formal solutions of
$\, L_4^{(4)}$, $z$ being the local variable of the expansion.
The singularities at the roots of $\,\, t^2\, -10\,t\, +1 =\,\,0\,$ are 
apparent.
This order-four operator (\ref{l44}) is actually homomorphic to its 
(formal) adjoint: 
\begin{eqnarray}
\label{intertwi}
adjoint(L_2) \cdot L_4^{(4)} \,\,\, \,\, = \,\, \,\,\,\,
 adjoint(L_4^{(4)}) \,   \cdot \, L_2, 
\end{eqnarray}
where  $\,L_2$ 
is the order-two intertwinner: 
\begin{eqnarray}
\label{L2}
&&L_2 \,\, = \, \,\, \Bigl(   D_t \, - \, 
{{d} \over {dt}} \ln(r(t)) \Bigr) \cdot D_t    \qquad   \quad \quad  \hbox{where:}
 \nonumber \\
&&r(t)  \, \, = \, \,\,  {{(t^2\, -10\, t\, +1) (t\, +1)} \over {t \cdot (t-1)^3}}.
\end{eqnarray}
The remarkable equivalence (\ref{intertwi}) of operator (\ref{l44}) with its adjoint 
is related to the fact that the exterior square of (\ref{l44}) 
has a rational function solution, 
that is, that this exterior square factors
 into an order-five operator $\, L_5$ and an  
order-one operator with a rational function solution (which coincides with
 $\, r(t)$ in (\ref{L2})).
\begin{eqnarray}
\label{ext2}
Ext^2(L_4^{(4)}) \, \, \, =\, \, \, \,\, L_5 \cdot 
\Bigl(D_t \, - \, {{d} \over {dt}} \ln(r(t))  \Bigr).
\end{eqnarray}
In other words, the (irreducible) order-four operator (\ref{l44}) 
is not only globally nilpotent ({\em ``Derived from Geometry''}~\cite{jm6}),
it is a ``special'' G-operator~\cite{Baldassarri} (Special Geometry):
its {\em differential Galois group} 
becomes ``special'' (symplectic or orthogonal groups, 
see for instance~\cite{Katz}).

This highly selected character of the  order-four operator (\ref{l44}) 
is  further confirmed by the {\em ``integrality property''}~\cite{jm9}
 of the series expansion 
of its analytical solution at $\, t \, = \, 0$:
\begin{eqnarray}
\label{sol}
&&Sol(L_4^{(4)}) \, \,  =\, \,\, \, 
t \, \,+11/8\, t^2\,+27/16\,t^3\,+2027/1024\,t^4\,
\nonumber \\
&& \qquad \qquad +9269/4096\,t^5\,
+83297/32768\,t^6\, + \, \, \cdots 
\end{eqnarray}
which, after rescaling $\, t \, = \, 16 \, u$, becomes 
a series with {\em integer} coefficients:
\begin{eqnarray}
&&\hspace{-0.5in}Sol(L_4^{(4)}) \,  \, =\, \, \,\,
 16\, u\, +352\, u^2\, +6912\, u^3\, 
+129728\, u^4\, +2372864\, u^5\,\nonumber \\
&&\hspace{-0.5in} \qquad  +42648064\, u^6\,
 +756609024\, u^7\, +13286784384\, u^8\, \\
&&\hspace{-0.5in} \qquad  +231412390144\, u^9\,
 +4002962189824\, u^{10}\, +68843688570880\, u^{11}\, 
\nonumber \\
&&\hspace{-0.5in}\qquad
+1178125203260416\, u^{12}\, +20074611461902336\, u^{13}\,
 \nonumber \\
&&\hspace{-0.5in} \qquad 
+340769765322760192\, u^{14}\, 
+5765304623564259328\, u^{15}\,\nonumber \\
&&\hspace{-0.5in}\qquad +97249731220784896768\, u^{16}\,
  +1636034439292348588288\, u^{17}\, 
 \, + \, \, \cdots \nonumber 
\end{eqnarray}
This {\em integrality property}~\cite{Kratten} suggests a 
{\em modularity}~\cite{jm9,SP4,LianYau} of this 
order-four operator (\ref{l44}). The simplest scenario would correspond
to (\ref{sol}) being elliptic integrals or, beyond,  modular forms
 that would typically be (up
to differential equivalence) a $\, _2F_1$ hypergeometric function with
{\em not one, but two pullbacks} (the relation between these 
two pullbacks being a modular curve).  More involved scenarios 
would correspond to {\em Calabi-Yau ODE's}~\cite{Almkvist,TablesCalabi}
 and some other {\em mirror maps} (see~\cite{jm9}).
We have first explored the simplest scenarios 
(elliptic integrals,  {\em modular forms}), which, as far as differential 
algebra is concerned, amounts to seeing if  this 
order-four operator (\ref{l44}) can be reduced, up to differential operator
equivalence, to symmetric powers of a second order operator.
This simple scenario is ruled out\footnote[1]{
See for instance, van Hoeij's program~\cite{ISSAC} from ISSAC'2007.}.
 We are now forced to explore 
the, much more complex, Calabi-Yau framework, with two 
possible scenarios: a general Calabi-Yau order-four
 ODE~\cite{Almkvist,TablesCalabi}, 
or  a Calabi-Yau order-four ODE that is $\, _4F_3$ solvable, the solution like 
(\ref{sol}) being expressed, up to operator equivalence, in term of 
a  $\, _4F_3$ hypergeometric function 
{\em up to a pullback that remains to be 
discovered}. This last situation would correspond
 to the $\, _4F_3$ Calabi-Yau 
situation we already encountered in
 $\, {\tilde \chi}^{(6)}$ (see~\cite{jm9}).
The $\, _4F_3$ solvability is clearly a desirable situation,
 because everything can be much more explicit. 

In constrast with the (globally nilpotent)
order-two operators,
finding that a given order-four operator corresponds 
to a given  $\, _4F_3$ operator up to a pullback 
(and up to homomorphisms) is an extremely difficult task, 
because the necessary techniques have not yet been developed.
Quite often, it goes the other way (no go result): assuming a rational pullback,
 one can rule out 
 a given order-four operator being 
a $\, _4F_3$ operator with a rational pullback (up to differential
operator equivalence).

In fact, and fortunately, operator (\ref{l44}) turns
 out to be,  
a nice example. It has singularities at $ \, 0,  \, 1, \,  -1, \, \infty$, and 
 these points have to be mapped  to
 $ \, 0,  \, 1,  \, \infty$ (i.e. the 
singularities of $\, _4F_3$ hypergeometric functions),
by the pullback.
{\em Assuming a rational pullback of degree two}, there is 
a systematic algorithm
to find all of the rational pullbacks of degree two
mapping $ \, 0,  \, 1, \,  -1, \, \infty$
 onto  $ \, 0,  \, 1,  \, \infty$.
This systematic algorithm is described in~\cite{systematic}
for order-two operators, but the same approach works (with little change)
 for fourth order operators as well.
The rational pullback function can actually be
 obtained\footnote[5]{For order-two equations 
with four singularities (HeunG ...), there are already hundreds of cases 
(now all found), see~\cite{overview}. Looking at the size of that
 table~\cite{overview} it is clear that providing an algorithm for 
finding pullbacks will be quite hard.} 
(with some trial and error) from this mapping of singularities
constraint and from the exponent-differences, in the same way as
 in section 2.6 in~\cite{BCHP}. The reader who is just interested in the 
surprisingly simple final result and not the mathematical structures, 
in particular the interesting relations between
some Calabi-Yau ODEs and selected $\, _4F_3$, can skip the next three
subsections\footnote[1]{Which correspond, in fact, to the 
way we originally found the result.} and jump directly to
 the solution of (\ref{l44})
given by (\ref{chi43}) with (\ref{A3}).  

\subsection{Simplification of $\, L_4^{(4)}$} 
\label{simplifL44}

As a ``warm up'', let us, for the moment, try to simplify the order-four operator
 (\ref{l44}), getting rid of the apparent singularities 
 $\,\, t^2\, -10\, t\, +1 =\,0$, and  trying to take into account all
the symmetries of (\ref{l44}): for instance, one easily remarks that
 (\ref{l44}) is actually invariant by the 
involutive symmetry $\, t \, \leftrightarrow \, 1/t$.

Let us introduce the order-four operator
\begin{eqnarray}
\label{l44t}
&&\hspace{-0.8in} {\mathcal L}_4 \,\, \, = \, \,\,   \, \,
D_x^4  \,
+ \,  {{10 \, x^2\,-2\,x \,-5} \over {(x-1)\,  (1 \, + \, 2 \, x) \, x }} 
  \cdot D_x^3  \, \, \, 
    +\, {{1} \over {4}} \cdot 
     {{ (5 x + 4) \cdot (6\, x^2\,-13\,x \,+4)  } \over 
 { (x-1)^2 \,  (1 \, + \, 2 \, x) \, x^2 }}   \cdot D_x^2 
\nonumber \\
&&\hspace{-0.8in} \quad \quad \quad    +\,  {{1} \over {4}} \cdot 
                 {{x \, +\, 8}
 \over { (x-1)^2 \,  (1 \, + \, 2 \, x) \, x^2 }}  \cdot D_x\, \, \, \,
    - \,{{3} \over {4 \cdot (x-1) \,  (1 \, + \, 2 \, x) \, x^3 }},
\end{eqnarray}
where $\, 1 \, + \, 2 \, x \, = \, \, 0$ is an apparent singularity. 
One can easily verify that the order-four operator (\ref{l44})
is the previous operator (\ref{l44t}), where we have 
performed the  $\, t \, \leftrightarrow \, 1/t$ invariant pullback:
\begin{eqnarray}
\label{pullback}
x \, \,  = \, \,  \, -\, {{4 \, t } \over {(1\, -t)^2 }}, \quad \qquad
L_4^{(4)} \, = \, \,  \,
 {\mathcal L}_4\Big[x \, \rightarrow \, -\, {{4 \, t } \over {(1\, -t)^2 }} \Bigr]. 
\end{eqnarray}

The operator (\ref{l44t}) is  
homomorphic to another order-four 
operator with {\em no apparent singularities}
\begin{eqnarray}
\label{l44other}
&&\hspace{-0.7in}{\mathcal M}_4 \,\, = \, \,  \,\, 
D_x^4 \, \,  + \, 2 \cdot  {{5\,x \,-4} \over {(x-1)\cdot \, x }} 
  \cdot D_x^3 \, 
    +\, {{1} \over {4}} \cdot 
     {{ (95\, x^2\,-160\,x \,+56)  } \over 
 { (x-1)^2 \cdot \, x^2 }}   \cdot D_x^2 
\nonumber \\
&&\hspace{-0.5in} \quad \quad   +\,  {{1} \over {4}} \cdot 
                 {{45\,x^3\,-124\,x^2\,+104\,x\,-16}
 \over { (x-1)^3 \cdot \, x^3 }}  \cdot D_x\, \, \,
    - \,{{2 \, x\, -\, 5} \over {4 \cdot (x-1)^3 \cdot \, x^3 }},
\end{eqnarray}
as can be seen by the (very simple) intertwinning relation:
\begin{eqnarray}
\label{homo}
 {\mathcal M}_4 \cdot D_x \,\,\, \, = \,\, \,  \, \, \,
\Bigl(D_x \, + {{ 10\,x^2\,-4\,x\,-3} \over {
(x-1)\,  (1\, + \, 2\, x) \cdot x }}\Bigr) \cdot  {\mathcal L}_4. 
\end{eqnarray}
This last operator with no apparent singularities, 
is homomorphic to its adjoint in a very simple
way:
\begin{eqnarray}
 adjoint({\mathcal M}_4) \cdot x^4 \cdot (1-x)
 \, \,\,\,\,  = \,\, \, \,  \, \, \,
x^4 \cdot (1-x) \cdot {\mathcal M}_4. 
\end{eqnarray}
{\em Do note that, remarkably, the exterior square of} $\, {\mathcal M}_4$, 
{\em  is an order five 
operator} and not the order six operator one could expect generically
from an intertwinning relation like (\ref{homo}) (the  exterior square of 
the order-four operator (\ref{l44t})
is order six with a rational function solution $\, (1\, +2\, x)/x$). 
Taking into account all these last results (no apparent singularities,
the singularities being the standard $\, 0$, $\, 1$, $\infty$
singularities, the intertwinning relation (\ref{homo}), the fact that
the exterior square is of order five), the order-four operator (\ref{l44other})
looks like a {\em much simpler} operator to study 
than the original operator (\ref{l44}). 

\subsection{$k$-balanced  $\,_4F_3$ hypergeometric function} 
\label{hypergeomRemark}

Let us make here an important preliminary remark on
the  $\, _4F_3$ linear differential operators. 
Let us consider a  $\, _4F_3$ hypergeometric function
\begin{eqnarray}
\label{4F3generic}
 _4F_3([a_1, \, a_2, \, a_3, \, a_4], \, [b_1, \, b_2, \, b_3]; t), 
\end{eqnarray}
with {\em rational values} of the parameters $\, a_i$ and $\, b_j$.
Its exponents at $\, x=\, 0$  are $\, 0$, $\,1 \, -\, b_1$,   $\,1 \, -\, b_2$,  
 $\,1 \, -\, b_3$,  its exponents at $\, x= \, \infty$
are  $\,a_1$,  $\,a_2$,  $\,a_3$,  $\,a_4$, and its exponents at 
$\, x\, = \, \, 1$ are  $\,0$,   $\,1$,  $\,2$ and $\, {\cal S}$
where  $\, {\cal S}$ is the
{\em  Saalsch{\"u}tzian
difference}:
\begin{eqnarray}
\label{diff}
 {\cal S} \,\,\, = \,\, \, \,  (b_1\, +\,  b_2 \, +\,  b_3)
 \,\, \, -(a_1\, +\,  a_2 \, +\,  a_3  +\,  a_4).
\end{eqnarray}
The  Saalsch{\"u}tzian
 condition~\cite{Saal,Saal2,Saal3} 
$\,  {\cal S} \, = \, \, 1$ is thus a  condition of confluence
 of two exponents at $\, x \, = \, \, 1$. 

The linear differential
 order-four operators annihilating the $\, _4F_3$ hypergeometric function 
(\ref{4F3generic})  are necessarily globally nilpotent,
and {\em they will remain globally nilpotent up to pullbacks
and up to differential operator 
equivalence}\footnote[2]{Global nilpotence is preserved
by pullbacks (change of variables) and by 
homomorphisms (operator equivalence).}. In contrast, 
the corresponding order-four operators 
are not, for generic (rational) values 
 of the parameters $\, a_i$ and $\, b_j$, such that they are 
homomorphic to their (formal) adjoint (``special geometry''), or such that
their exterior square of order-six
has a rational function solution (a degenerate
 case corresponding to the exterior square 
being of order {\em five}).

These last ``special geometry'' conditions
 (see (\ref{intertwi}) and (\ref{ext2})), correspond to 
selected algebraic subvarieties 
in the parameters $\, a_i$ and $\, b_j$.
 In the particular case of the
 {\em exterior square of the order-four operator
being of order five}\footnote[1]{This condition is seen by 
some authors, see (11) in~\cite{fast},
as a condition for the ODE to be a {\em Picard-Fuchs equation}
of a Calabi-Yau manifold. These conditions, namely (11) in~\cite{fast},
are preserved by pullbacks, not operator equivalence.},  we will show, 
in forthcoming publications,
that the parameters  $\, a_i$ and $\, b_j$
of the hypergeometric functions
are necessarily restricted to three sets of 
algebraic varieties:  a {\em codimension-three
algebraic variety} included in the  
Saalsch{\"u}tzian condition~\cite{Saal,Saal2,Saal3} 
$\,  {\cal S} \, = \, \, 1$ and two (self-dual for the adjoint)
{\em codimension-four algebraic varieties},
respectively included in the two hyperplanes 
$\,  {\cal S} \, = \, \, -1$ and
 $\,  {\cal S} \, = \, \, 3$.  

Imagine that one is lucky enough to see the order-four operator
(\ref{l44other}) (which is such that its  exterior square 
is of order five) as a $\, _4F_3$ solvable
Calabi-Yau situation: one is, thus, exploring 
particular  $\, _4F_3$ hypergeometric functions corresponding to
these (narrow sets of) algebraic varieties which
 single out,  particular ($\, k\, = \, -1, \, 1, \, 3$)
 {\em $\, k$-balanced} hypergeometric
 functions\footnote[5]{$\, k$-balanced hypergeometric functions
correspond to the Saalsch{\"u}tzian difference  (\ref{diff}) being 
an integer $k$: $\,  {\cal S} \, = \, \, k$.}
 (rather than the well-poised 
hypergeometric functions, or very well-poised\footnote[8]{Note that
very well-poised hypergeometric series are known~\cite{Rivoal}
to be related with $\, \zeta(2)$,$\, \zeta(3)$, ...,  which are
constants known to occur in the Ising model~\cite{jm4}.} 
hypergeometric functions~\cite{Mishev2,Mishev} one could 
have imagined\footnote[1]{Note that the conditions to be
well-poised hypergeometric series are actually preserved 
by the transformation 
$\, a_i \, \rightarrow \, \, 1\, -a_i, \, $
$b_j \, \rightarrow \, \, 2\, -b_j$, which corresponds to 
changing the linear differential operator, associated with 
hypergeometric functions, into its formal adjoint.}). We are actually
working up to operator equivalence (which amounts to performing 
derivatives of these hypergeometric functions). It is straightforward to
see that the $\, n$-th derivative
of a hypergeometric function shifts the Saalsch{\"u}tzian difference (\ref{diff}) 
by an integer, and that this  does not
preserve the condition for the exterior square of the 
corresponding order-four operator to be order five: it
 becomes an order-six operator (homomorphic to its formal adjoint) with 
a rational function solution. The natural framework for seeking 
  $\, _4F_3$ hypergeometric functions  (if any) for our order-four operators 
(\ref{l44}), (\ref{l44t}), (\ref{l44other}) 
is thus (selected) $\, k$-balanced hypergeometric functions (rather than the 
well-poised, or very well-poised, hypergeometric 
functions~\cite{Mishev2} ...). 

\subsection{$\, L_4^{(4)}$ is $\,_4F_3$ solvable, up to a pullback} 
\label{4F3solvable}

Let us restrict ourselves to the, at first sight simpler,  
order-four operators (\ref{l44t}), (\ref{l44other}): 
even if we know exactly 
the rational values of the parameters $\, a_i$ and $\, b_j$,
finding that a given order-four operator corresponds 
to this given  $\, _4F_3$ operator, up to a pullback 
(and up to homomorphisms), remains a quite difficult task.
We have first studied the case where the pullback 
in our selected  $\, _4F_3$ hypergeometric functions 
is a rational function. This first scenario has been 
ruled out on arguments based on the matching of the singularities
and of the exponents of the singularities.

We thus need to start exploring pullbacks that are 
 {\em algebraic functions}. Algebraic functions
 can branch at certain points (this can, for instance,
 turn a regular point into a singular point).
The set of algebraic functions is a very large one, so we 
started\footnote[2]{And also because we had a Ising model prejudice 
in favour of square roots~\cite{2008-experimental-mathematics-chi} ...} 
with the simplest algebraic functions 
situation namely, {\em square roots singularities}. A first
examination of the  matching of the singularities, and 
of the exponents of the singularities, indicates that we should 
have square roots at $\, x \, = \, \, 1$ only. 

Along this square root line, let us recall
 the well-known {\em inverse Landen transformation}
 in terms of $\, k$, the {\em modulus of the elliptic functions 
parametrizing the Ising model}:
\begin{eqnarray}
k\, \quad \longrightarrow \qquad 
{{ 1\, - \, \, \sqrt{1\, -k^2} } \over { 1\, + \, \, \sqrt{1\, -k^2} }}. 
\end{eqnarray}

In terms of the  variable $\, x \, = \, \, k^2$, this 
 inverse Landen transformation reads:
\begin{eqnarray}
\label{Pull}
&&\hspace{-0.5in}x \, \quad \longrightarrow \qquad 
P(x) \, \, \,\, = \, \,\, \,\,  \, 
\Bigl({{ 1\, - \, \, \sqrt{1\, -x} } \over { 1\, + \, \, \sqrt{1\, -x} }}\Bigr)^2 
\nonumber \\
&&\quad \quad \quad \quad \, \, = \, \,\,\,  \,  \,
  {{x^2 \, -8 \, x \, +8} \over {x^2}} \,\, \, \,
 -4 \cdot (2-x) \cdot {{(1-x)^{1/2} } \over {x^2}}.
\end{eqnarray}
Using this pullback $\, P(x)$, we have  actually been able to 
obtain the solution of the order-four differential operator (\ref{l44other}) 
 in terms of four terms like 
\begin{eqnarray}
\, \, _4F_3\Bigl(
[ {{1} \over {2}}, \, {{1} \over {2}}, \, {{1} \over {2}}, \,  
{{1} \over {2}}], \, [1, \, 1, \, 1]; \, P(x)  
\Bigr).  
\end{eqnarray}
This slightly involved solution is given
 in \ref{M4}.

We can now get the solution of (\ref{l44}), the original
operator $\, L_4^{(4)}$, 
from this slightly involved result, since (\ref{l44})
is (\ref{l44other}) up to a simple pullback
namely the change of variable (\ref{pullback}).
Coming back with (\ref{pullback}), to the original variable
 $\, t$ in $\, L_4^{(4)}$, 
the previous pullback (\ref{Pull})  simplifies remarkably: 
\begin{eqnarray}
\label{here}
P\Bigl( -\, {{4 \, t } \over {(1\, -t)^2 }} \Bigr)
 \, \,\, = \,\, \, \, \, {{1 \,+\,  t^4 } \over {2 \cdot t^2 }} 
\,\, -\,  {{1 \,-\,  t^4 } \over {2 \cdot t^2 }} 
\, \,\, = \, \, \, \, t^2, 
\end{eqnarray}
the Galois conjugate of (\ref{Pull}) giving $\, 1/t^2$. Of course,
 once this key result is known, namely that
 a $\, t^2$ pullback works,
it is easy to justify, 
a posteriori, this simple monomial result: after all, $\, L_4^{(4)}$
 has singularities at $ \, 0,  \, 1, \,  -1, \, \infty$, and 
these points can be mapped (under $\,t^2$) to
 $ \, 0,  \, 1,  \, \infty$ (i.e. the 
singularities of $\, _4F_3$ hypergeometric functions). 

Pullbacks have a natural structure with respect to {\em composition
of functions}\footnote[1]{Suppose that an operator $O_2$ is a pullback of 
an operator $O_1$,  where the pullback $f$ is
a rational function and that $O_3$ is also a pullback of $O_1$,
 where the pullback is a rational function $\,g$. Then $O_3$ is
 also a pullback of $O_2$.  To compute this pullback
function,  one has to compose $\,g$ and the inverse of $f$.}.
It is worth noting that (\ref{here}) describes the {\em composition
of two well-known isogenies of elliptic curves}, the {\em inverse Landen
 transformation} (\ref{Pull}), and the {\em rational isogeny} $\, -4t/(1-t)^2$
underlined by R. Vidunas~\cite{Vidunas2011} and in~\cite{Renorm}, 
giving the simple quadratic transformation
 $\, t \, \rightarrow \, t^2$.

All this means that the solution of $\, L_4^{(4)}$ can be expressed in terms of 
the hypergeometric function 
\begin{eqnarray}
\label{4F3t2}
 _4F_3\Bigl(
[ {{1} \over {2}}, \, {{1} \over {2}}, \, {{1} \over {2}}, \,  {{1} \over {2}}], \, 
[1, \, 1, \, 1]; \, t^2  \Bigr),   
\end{eqnarray}
and its derivatives.
Actually considering the hypergeometric operator $\,  {\mathcal H}$ 
having (\ref{4F3t2})
as a solution, it can be seen to be homomorphic to (\ref{l44})
\begin{eqnarray}
\label{interAA}
{\mathcal A}_3 \cdot  {\mathcal H} 
  \,\, \, \,= \, \, \, \, \,\, L^{(4)}_{4} \cdot A_3, 
\end{eqnarray}
where  the order-three intertwinners $ \, A_3$ and
  $\,{\mathcal A}_3$ read, respectively,
(with $\, d_3(t) \, = \, \,$ $ t \cdot (t+1) \cdot (t-1)^2 \cdot (t^2-10\,t+1)$): 
\begin{eqnarray}
\label{A3}
&&A_3 \,  \,\, = \, \,\,  \,  2 \cdot  (1+t) \cdot t^3 \cdot D_t^3 \,  \,
+\frac{2}{3} \cdot \frac{16t^2-t-11}{t-1} \cdot t^2 \cdot D_t^2
\nonumber \\
&&\qquad \quad 
+\frac{1}{3} \cdot \frac{31t^2-4t-11}{t-1}\cdot t  \cdot D_t \,\,  +  \, t, 
\end{eqnarray}
\begin{eqnarray}
\label{AA3}
&&\hspace{-0.5in}{\mathcal A}_3 \,\,  = \, \, \, \,  {{2} \over {t\, -1}} \cdot D_t^3
\, \,  \, 
+ {{2} \over {3}} \cdot  {{1} \over {d_3(t)}} \cdot
  (10\,t^4-107\,t^3-225\,t^2 +163\,t -17) \cdot D_t^2
\nonumber \\
&&\hspace{-0.5in} \quad \quad \quad 
+ {{1} \over {3}} \cdot 
 {{1} \over {t\cdot d_3(t)}} \cdot \,(5\,t^4-66\,t^3-900\,t^2 +290\,t -33) \cdot D_t
\nonumber \\
&&\hspace{-0.5in} \quad \quad \quad 
+ \, {{1} \over {3}} \cdot 
{{1} \over {t\cdot d_3(t)}} \cdot \,(t^3-21\,t^2\,+99\,t-23).
\end{eqnarray}

From the intertwinning relation (\ref{interAA}),
one easily finds that 
the solution of $\, L^{(4)}_{4}$ which is analytic
 at $\, t\, = \, \, 0$, is $\, A_3$ acting on (\ref{4F3t2}):
\begin{equation}
\label{chi43}
{\tilde \chi}^{(4)}_{d;3}(t) \,\,\, =   \, \,\, \,\,
A_3 \Bigl[
{}_4F_3([1/2, \,1/2, \,1/2, \,1/2],[1, \,1, \,1]; \, t^2)\Bigr].
\end{equation}

Having with (\ref{chi43}) a normalization for $\, {\tilde \chi}^{(4)}_{d;3}$,
we can now fix the values  of the coefficients
 $a_i$ in the sum (\ref{chisumL8}) for
${\tilde \chi}^{(4)}_{d}(t)$. They can be fixed 
by expanding and matching the rhs of (\ref{chisumL8}) 
and ${\tilde \chi}^{(4)}_{d}(t)$. This gives
\begin{eqnarray}
\label{chi4exp}
{\tilde \chi}^{(4)}_{d}(t)\,\,\,  =\, \, \,\,\,\, 
\frac{1}{2^3}\cdot  {\tilde \chi}^{(4)}_{d;1}(t)
\,\,\, +\frac{1}{3\cdot 2^3}\cdot {\tilde \chi}^{(4)}_{d;2}(t)
\,\,\, -\frac{1}{2^3}\cdot {\tilde \chi}^{(4)}_{d;3}(t).
\end{eqnarray}
\vskip .2cm 

{\bf Remark:} It is quite surprising to find exactly the 
same $_4F_3$ hypergeometric function (\ref{4F3t2}) with  the exact {\em same}
remarkably simple pullback $\, t^2$, as  the one we already found~\cite{jm9} 
in the order-four Calabi-Yau operator $\, L_4$ in $\, {\tilde \chi}^{(6)}$.
\vskip .2cm 

{\bf Comment:} Of course, from a mathematical viewpoint, when looking for a pullback, 
one can in principle always ignore all apparent singularities. 
These calculations displayed here look a bit paradoxical: the calculations
performed with the (no apparent singularities)
operator (\ref{l44other}), which  looks {\em simpler} 
(it has an exterior square of order five,
 and is very simply homomorphic to its adjoint, ...)
turns out to have a more complicated pullback (\ref{Pull}), than the 
amazingly simple pullback (namely $\, t^2$)
we finally discover for the original operator (\ref{l44}) (see (\ref{chi43})).
The ``complexity'' of the original operator (\ref{l44}) 
is mostly encapsulated in the order-three intertwinner $ \, A_3$  
(see (\ref{A3})). The ``$_4F_3$-solving'' of the operator amounts to 
reducing the operator, up to operator equivalence (\ref{interAA}),
to a $_4F_3$ hypergeometric operator up to a pullback. Finding the pullback is 
the difficult step: as far as ``$_4F_3$-solving'' of an operator is concerned,
what matters is the {\em complexity of the pullback}, not the 
 complexity of the operator equivalence.
\vskip .2cm 

{\bf Ansatz:} Of course, knowing the key ingredient in the final result 
(\ref{chi43}), namely that the pullback is just $\, t^2$, it would have 
been much easier to get this result. Along this line one may recall the conjectured
existence of a natural boundary at unit circle $\, |t| \, = \, \, 1$ 
for the full susceptibility of the Ising model, and, more specifically
 for the diagonal susceptibility $\, n$-fold integrals we study here, 
the fact that the singularities are all $\, N$-th roots of unity ($\, N$  integer).
Consequently, one may have, for the Ising model,
 a  $\, t^N$ ($N$  integer) prejudice for pullbacks.

In forthcoming studies of linear differential operators occurring
 in the next (bulk) $\, {\tilde \chi}^{(n)}$'s or (diagonal) 
$\, {\tilde \chi}^{(n)}_d$'s, when trying to  see if these new 
(Calabi-Yau like, special geometry) operators are $\, _{q+1}F_q$
 reducible up to a pullback, we may save
some large amount of work  by assuming that the corresponding pullbacks are 
of the simple form $\, t^N$ where $\, N$ is an integer.

\section{The linear differential equation of ${\tilde \chi}^{(5)}_d$ in mod. prime
and exact arithmetics}
\label{chi5}
The first terms of the series expansion of $\, {\tilde \chi}_d^{(5)}(x)$ read:
\begin{eqnarray}
\label{first}
&&\hspace{-0.5in} {\tilde \chi}_d^{(5)}(x)
\, \, = \, \,\, \, {{3} \over {262144}} \cdot x^{12}
\, \,+ \,{{39} \over {1048576}} \cdot x^{14}
\,\, + \,{{5085} \over {67108864}} \cdot x^{16} \, \\
&&\hspace{-0.5in} \quad \quad \quad + \, {{9} \over {67108864}} \cdot x^{17}\,
+ \,{{33405} \over {268435456}} \cdot x^{18} \,
+ \,{{ 315} \over {536870912}} \cdot x^{19} \, + \, \cdots \nonumber
\end{eqnarray}
where $\,x=\, t^{1/2}\, =\, \sinh2E_v/kT \, \sinh 2E_h/kT$ is our independent variable.
In order to obtain the linear differential equation for $\, {\tilde \chi}_d^{(5)}(x)$,
we have used in \cite{mccoy3} a ``mod. prime'' calculation
which amounts to generating large series {\em modulo
a given prime}, and then deduce, the linear differential operator for
$\, {\tilde \chi}_d^{(5)}(x)$ {\em modulo that prime}.
With $\,3000 $ coefficients for the series expansion of $\, {\tilde \chi}_d^{(5)}(x)$
modulo a prime, we have obtained linear
differential equations of order $\, 25,\, 26, \, \cdots$.
The smallest order we have reached is $\, 19$, and we have {\em assumed}
 that the linear
differential equation of $\, {\tilde \chi}_d^{(5)}(x)$ is of minimal order 19.
In~\cite{2008-experimental-mathematics-chi}, we have introduced
a method to obtain the minimal order of the ODE by producing some
$( \ge 4)$ non minimal order ODEs and then using the "ODE formula"
(see~\cite{jm7,jm8,2008-experimental-mathematics-chi} for details and how to
read the ODE formula). The ODE formula for ${\tilde \chi}_d^{(5)}(x)$ reads
\begin{eqnarray}
31\, Q\, + 19 \, D\, -302 \,\,\,=\,\,\,\, \,(Q+1) \cdot (D+1)\,\, -f,
\end{eqnarray}
confirming that the minimal order of the ODE for $\, {\tilde \chi}_d^{(5)}(x)$ is 19.
Note that the degree of the polynomial carrying apparent singularities should
be 237 (see Appendix B in~\cite{jm7}).
Call $\, {\cal L}^{(5)}_{19}$ the differential operator (known mod. prime)
for $ \, {\tilde \chi}_d^{(5)}(x)$.
The singularities and local exponents of
$\, {\cal L}^{(5)}_{19}$ are\footnote{
The local exponents are given as (e.g.) $2^3$ meaning $2, 2, 2$.}
\begin{eqnarray}
\label{L719}
x=0,\qquad && \rho = \, 0^5, 1/2, 1^4, 2^3, 4^3, 3, 7, 12,
\nonumber \\
x=\infty, \qquad && \rho = \, 1^5, 3/2, 2^4, 3^3, 4, 5^3, 8, 13,
\nonumber \\
x=1,\qquad && \rho = \, -3, -2, -1, 0^4, 2^3, 4^2,\, \cdots,
\\
x=-1,\qquad && \rho = \, 0^5, 2^4, 4^3, 6^2, 8^2, 10^2,\, \cdots,
\nonumber \\
x=x_0, \qquad && \rho = \, 5/2, 7/2, 7/2,\, \cdots,
\nonumber \\
x=x_1,\qquad && \rho = \, 23/2, \, \cdots
\nonumber
\end{eqnarray}
where $x_0$ (resp. $x_1$) is any root
of $\, 1+x+x^2=\, 0$ (resp. $1+x+x^2+x^3+x^4=\, 0$),
and the trailing $\cdots$ denotes integers not in the list.
Note that, in practice, we do not deal with the minimal order differential
operator $\, {\cal L}^{(5)}_{19}$ but with an operator of order 30
(that $\, {\cal L}^{(5)}_{19}$ rightdivides): order 30 is what we have called
in~\cite{jm7,jm8,2008-experimental-mathematics-chi}
the ``optimal order'', namely the order for which
finding the differential operator annihilating the series requires
the minimum number of terms in the series.
With the tools and methods developed
in~\cite{jm7,jm8,2010-chi5-exact}, we are now able
to factorize the differential operator and
{\em recognize some factors in exact arithmetic}.
This way, we may see whether some factors occurring $\, {\cal L}^{(5)}_{19}$
follow the "special geometry" line we encountered for
$\, {\tilde \chi}_d^{(3)}(x)$ and $\, {\tilde \chi}_d^{(4)}(t)$.
Our first step in the factorization of $\, {\cal L}^{(5)}_{19}$ is to check
whether $\, {\cal L}^{(3)}_{6}$ (the differential
operator for $ \, {\tilde \chi}_d^{(3)}$)
is a right factor of $\, {\cal L}^{(5)}_{19}$, meaning
that the solutions of $\, {\cal L}^{(3)}_{6}$
(and in particular the integral $ \, {\tilde \chi}_d^{(3)}(x)$)
are also solution of $\, {\cal L}^{(5)}_{19}$.
This is indeed the case.
Using the methods developed in~\cite{jm7,jm8,2010-chi5-exact}, we find
that the series for the difference
$ \,{\tilde \chi}_d^{(5)}(x) \, - \alpha \, {\tilde \chi}_d^{(3)}(x)$
requires an ODE of minimal order 17 for the value\footnote[5]{
Comparing with eq.(58) in~\cite{mccoy3}, one should not expect a
$\, 1/2$ contribution, since the sum on the $\, g^5(N,t)$'s still
contains $\tilde{\chi}^{(3)}_d$.}
$ \,\alpha \,= \, 8$. This
confirms that $\, {\cal L}_6^{(3)}$
is in direct sum in $\, {\cal L}^{(5)}_{19}$, and that some
(order-four) factors of $\,{\cal L}_6^{(3)}$ are {\em still}
in $\,{\cal L}_{17}^{(5)}$:
\begin{eqnarray}
{\cal L}^{(5)}_{19} \,\,\, = \,\,\, \,
{\cal L}^{(3)}_{6} \,\oplus\, {\cal L}^{(5)}_{17}
\end{eqnarray}
This order four factor is obviously $\, L^{(3)}_1 \oplus L^{(3)}_3$.
Since these factors are in direct sum in $\, {\cal L}_6^{(3)}$,
the order-seventeen operator ${\cal L}^{(5)}_{17}$ is also the annihilator of
${\tilde \chi}_d^{(5)}(x) - \beta \, {\tilde \chi}_{d,2}^{(3)}(x)$
for $\,\beta \, = \, \, 4$, meaning that we also have
\begin{eqnarray}
\, {\cal L}^{(5)}_{19} \,\,\, = \,\,\, \, L^{(3)}_{2} \,\oplus\, {\cal L}^{(5)}_{17}
\end{eqnarray}
At this step, the differential operator ${\cal L}^{(5)}_{17}$
 is known in prime.
To go further in the factorization, we use the method
 developed in~\cite{jm7,jm8}
along various singularities and local
exponents of ${\cal L}^{(5)}_{17}$ which
read\footnote[1]{There are solutions analytic
at $x=\, 0$ with exponents $\,0,\,1,\,2,\,\,4,\,7$.
}:
\begin{eqnarray}
\label{L517}
\hspace{-0.5in}x=0, \qquad && \rho = \, 0^5, 1/2, 1^4, 2^3, 4^2, 3, 7,
\qquad \, \, \,\ln(z)^4,\,z^{1/2},
\nonumber \\
\hspace{-0.5in}x=\infty,\qquad && \rho = \, 1^5, 3/2, 2^4, 3^3, 4, 5^2, 8,
\qquad \,\, \,\ln(z)^4,\,\,z^{3/2},
\nonumber \\
\hspace{-0.5in}x=1, \qquad && \rho = \, -3, -2, -1, 0^4, 2^3, \cdots,
\qquad \ln(z)^3,\,z^{-3}, z^{-2}, z^{-1},
\nonumber \\
\hspace{-0.5in}x=-1,\qquad && \rho = \, 0^5, 2^4, 4^3, 6^2, 8^2, \cdots,
\qquad \quad \ln(z)^4,
\\
\hspace{-0.5in}x=x_0,\qquad && \rho = \, 5/2, 7/2, 7/2, \cdots,
\qquad \quad \,\,\,\,\,\,\, z^{5/2}, z^{7/2}, z^{7/2}\ln(z),
\nonumber \\
\hspace{-0.5in}x=x_1,\qquad && \rho = \, 23/2, \cdots,
\qquad \qquad \qquad \,\,\,\, z^{23/2}
\nonumber
\end{eqnarray}
where $x_0$ and $x_1$ are again the roots of $1+x+x^2=\, 0$
and $1+x+x^2+x^3+x^4= \, 0$, and the trailing $\cdots$ 
denotes integers not in the
list. The last column shows the maximum $\ln(z)$-degree
 occurring in the formal solutions of
$ \, {\cal L}^{(5)}_{17}$, $z$ being the local variable of the expansion.
 
Use is made of section 5 of~\cite{mccoy3} to recognize exactly some factors.
This is completed by an usual 
{\em rational reconstruction}~\cite{2010-chi5-exact}.
We are now able to give new results completing what
was given in section 5 of~\cite{mccoy3}.
The linear differential operator ${\cal L}_{17}^{(5)}$ has the factorization:
\begin{eqnarray}
\label{factoL17}
\hspace{-0.4in}{\cal L}^{(5)}_{17} \,\, =\,\,\, L^{(5)}_{6} \cdot {\cal L}^{(5)}_{11}.
\end{eqnarray}
The linear differential operator $\, {\cal L}^{(5)}_{11}$ has been fully factorized
and the factors {\em are known in exact arithmetic} (the indices are the orders)
\begin{eqnarray}
\label{factoL11}
\hspace{-0.4in}{\cal L}^{(5)}_{11} \, \,=\,\,\, \,
L^{(3)}_{1} \, \oplus \, L^{(3)}_{3} \, \oplus \,
\Bigl(W^{(5)}_{1} \cdot U^{(5)}_{1}\Bigr) \,
\oplus \, \Bigl(L^{(5)}_{4} \cdot V^{(5)}_{1} \cdot U^{(5)}_{1}\Bigr),
\end{eqnarray}
and are given in \ref{appL11}.
The factor $\, L^{(5)}_{6}$ is the only one which is known in
primes\footnote{This factor known only in primes
does not allow the computation of the singular behavior of $\tilde{\chi}^{(5)}_d$.}
and it is irreducible.
The irreducibility has been proven with the method presented in section 4
of~\cite{jm7}. This is technically tractable since there are only
two free coefficients (see (\ref{seriesL6}) below)
that survive in the expansion of the analytical
series at $x=\, 0$ of $L^{(5)}_{6}$.
In the factorization (\ref{factoL17}), (\ref{factoL11})
of ${\cal L}^{(5)}_{11}$ and ${\cal L}^{(5)}_{17}$, the
factors are either known and occur elsewhere ($L^{(3)}_{1}$, $L^{(3)}_{3}$)
or simple order-one linear differential operators
($U^{(5)}_{1}$, $V^{(5)}_{1}$, $W^{(5)}_{1}$), {\em except} the
order-four operator $L^{(5)}_{4}$ and the order-six operator $L^{(5)}_{6}$.
It is then for these specific operators that we examine whether they are
``Special Geometry''.

\subsection{The linear differential operator $L^{(5)}_{4}$}
\label{diffl54}
The order-four linear differential operator
$L^{(5)}_{4}$ has the following local exponents
\begin{eqnarray}
\label{localexp}
x\,=\,0, \qquad \quad&& \rho = -2, -2, -1,\, 0,
\nonumber \\
x\,=\,\infty, \qquad \quad&& \rho =\, 3,\, 3, \,4,\, 5,
\nonumber \\
x\,=\,1, \qquad \quad&& \rho = -2, -2, -2, -2,
\\
x\,=\,-1, \qquad \quad&& \rho = -2, -2,\, 0,\, 0,
\nonumber \\
1\,+x\,+x^2\,=\, 0, \qquad \quad&& \rho = -1, 0, 1, 2.
\nonumber
\end{eqnarray}
At all these singularities $x_0$, the solutions have the maximum allowed degree
of log's (i.e. $\ln(x-x_0)^3$), except at the singularities roots
of $\,1\,+x\,+x^2\,=\, 0$, where the solutions carry no log's.
In view of the negative local exponents in (\ref{localexp}),
 we introduce:
\begin{eqnarray}
\mu(x) \,\,\,=\,\,\, 
\Bigl(8 \cdot (x-1)^2 \cdot(x+1)^2 \cdot(x^2 \, +x\, +1)\Bigr)^{-1}.
\end{eqnarray}
Then, if we consider the linear differential operator
 $\mu(x)^{-1}\,  \cdot\,  L^{(5)}_{4} \, \cdot \, \mu(x)$,
nothing prevents (as far as the $\rho$'s
and log's are concerned) to check whether
this conjugated operator is homomorphic to a symmetric cube
of the order-two linear differential
operator of an elliptic integral.
We find the solution of $\,L^{(5)}_{4}$, which is analytic at $\,x\,=\, 0$,
 as a cubic expression of
 $_2F_1$ hypergeometric functions, with {\em palindromic} polynomials
\begin{eqnarray}
\hspace{-0.8in}&&Sol( L^{(5)}_{4} ) \, \,=\,\,\, \, 
\mu(x) \cdot \Bigl(3\, x^5 \cdot F_1^3 \, + \, 
+ \, 8 \cdot (2 \,  +2\,x +x^2 \, +2\,x^3 +2\,x^4) \cdot F_0^3
\nonumber \\
\hspace{-0.8in}&& \qquad \qquad \qquad 
+\, 4\, x\cdot (5 \, +x \, -3\, x^2 \, +x^3 \, +5\, x^4) \cdot F_1 \cdot F_0^2 
\nonumber \\
\hspace{-0.8in}&& \qquad \qquad \qquad 
+\, 2\, x^2 \cdot (2 \, -10\, x \, -17\, x^2 \, -10\, x^3 \, +2\, x^4) 
\cdot F_1^2 \cdot F_0  \Bigr),
\end{eqnarray}
where $\, F_0$ and $\, F_1$ are respectively ${}_2F_1([1/2, 1/2],[1]; x^2)$ 
and ${}_2F_1([1/2, 3/2],[2]; x^2)$
(closely related, up to a $\pi/2$ factor,
 to the usual complete elliptic integrals).
Again the occurrence of (very simple) elliptic integrals is underlined.
Note that $L^{(5)}_{4}$ contributes to the solutions of ${\cal L}^{(5)}_{17}$
in the block $L^{(5)}_{4} \cdot V^{(5)}_{1} \cdot U^{(5)}_{1}$ which has
the local exponents
\begin{eqnarray}
x\,=\,0, \qquad \quad&& \rho = \, 0^3,\, 1^2,\, 2,
\nonumber \\
x\,=\,\infty, \qquad \quad&& \rho =\, 1^3,\, 2^2, \,3,
\nonumber \\
x\,=\,1, \qquad \quad&& \rho = -3, -2,\, 0^4,
\\
x\,=\,-1, \qquad \quad&& \rho =\, 0^4,\, 2^2,
\nonumber \\
1\,+x\,+x^2\,=\, 0, \qquad \quad&& \rho = 0, 1, 2, 3, 4,\, 5/2.
\nonumber
\end{eqnarray}
There are two solutions analytic at $x=0$ with exponents 1 and 2.

\subsection{On the order-six linear differential
 operator $L^{(5)}_{6}$: ``Special Geometry'' }

Let us write the formal solutions  $L^{(5)}_{6}$ at $x=0$, 
where the notation $[x^p]$ means that the series begins 
as $x^p \,(const. + \cdots)$.
There is one set of five solutions and one extra
 solution analytical at $\, x \, = \, 0$ 
(i.e. four solutions with a log, and two 
 solutions analytical at $\, x \, = \, 0$):
\begin{eqnarray}
\label{seriesL6}
&& S_1 = [x^7] \, \ln(x)^4\, + [x^4] \, \ln(x)^3 \,
+[x^2] \, \ln(x)^2 \, + [x^0] \, \ln(x) \, + [x^0],  \, 
 \nonumber \\
&& S_2 = [x^7] \, \ln(x)^3 \,+
[x^4] \, \ln(x)^2 \, + [x^2] \, \ln(x) \, + [x^0],  \, 
 \nonumber \\
&& S_3 = [x^7] \, \ln(x)^2 \, + [x^4] \, \ln(x) \,+ [x^3],  \,  \\
&& S_4 = [x^7] \, \ln(x) \, + [x^4],  
\, \quad \quad \quad S_5 = [x^7],
 \qquad \quad \hbox{and:} \nonumber  \\
&& S_6 = [x^2]. \nonumber 
\end{eqnarray}
In view of this structure, the linear
 differential operator $L^{(5)}_{6}$ cannot be (homomorphic to)
a symmetric fifth power of the linear
 differential operator corresponding to the
elliptic integral.

The next step is to see whether the exterior square 
of $\,L^{(5)}_{6}$ has a rational solution,
which means that $\, L^{(5)}_{6}$ corresponds to "Special Geometry".
With the six solutions (\ref{seriesL6}), seen as series
 obtained mod. primes, one can 
easily built the general solution of $\, Ext^2 (L^{(5)}_{6})$ as
\begin{eqnarray}
\sum_{k,p} \, d_{k,p}\cdot  ( S_k \, {\frac{d S_p}{dx}} \, 
- S_p \, {\frac{d S_k}{dx}} ), \qquad k \ne p\, =\,\,1,\, \cdots,\,  6,  
\end{eqnarray}
which should not contain log's, fixing then some of the coefficients $d_{k,p}$.

For a rational solution of $\, Ext^2 (L^{(5)}_{6})$
 to exist, the form (free of log's)
\begin{eqnarray}
D(x) \cdot \, \sum_{k,p} \, d_{k,p}\cdot
  ( S_k \, {\frac{d S_p}{dx}}  \, - S_p \, {\frac{d S_k}{dx}}), 
\end{eqnarray}
should be a polynomial, where the denominator $\, D(x)$ reads
\begin{eqnarray}
\hspace{-0.8in} D(x)\, \,=\,\,\,\,
  x^{n_1} \cdot  (x+1)^{n_2}  \cdot (x-1)^{n_3} 
 \cdot (1+x+x^2)^{n_4}  \cdot (1+x+x^2+x^3+x^4)^{n_5},
 \nonumber
\end{eqnarray}
the order of magnitude of the exponents 
$\, n_j$ being obtained from the local exponents of
the singularities.
With series of length 700, we have found no rational 
solution for $\, Ext^2 (L^{(5)}_{6})$.

Even if $\, L^{(5)}_{6}$
 is an irreducible operator of {\em even} order,
we have looked for a rational solution for its {\em symmetric square}.
The general solution of $\, Sym^2 (L^{(5)}_{6})$
 is built from (\ref{seriesL6}) as
\begin{eqnarray}
\sum_{k,p} \, f_{k,p}\cdot  S_k \, S_p, 
 \qquad \quad k \,\ge\, p \,=\,1,\, \cdots,\, 6,  
\end{eqnarray}
and the same calculations are performed.
With some 300 terms, we actually found that $\, Sym^2 (L^{(5)}_{6})$ 
{\em has a rational solution} of the
form\footnote[1]{Note that this form  occurs, for a non minimal 
representative of $\, L^{(5)}_{6}$, in the 
factorization (\ref{factoL17}). 
On this point, see the details around (43), (44) in~\cite{jm7}.}
 (with $\, P_{196}(x)$ a polynomial of degree 196):
\begin{eqnarray}
\hspace{-0.4in} {\frac{x^4 \cdot  P_{196}(x)}{
(x+1)^{10} \cdot  (x-1)^{14} \cdot  (1+x+x^2)^{21}\cdot   (1+x+x^2+x^3+x^4)^{9}}}, 
\end{eqnarray}
thus showing that $\, L^{(5)}_{6}$ does correspond to "Special Geometry".

Note that the occurrence (\ref{seriesL6})
 of {\em two analytic} solutions at $x\, =\,\,  0$,
 for $\, L^{(5)}_6$, which is irreducible,
is a situation we have encountered
 in Ising integrals~\cite{jm7,2010-chi5-exact}.
The order twelve differential operator (called
 $\, L^{left}_{12}$ in~\cite{2010-chi5-exact}) has four analytical
solutions at $\, x\, =\, 0$ and it has been
 demonstrated that it is irreducible~\cite{2010-chi5-exact}.

\section{Singular behavior of ${\tilde \chi}^{(3)}_{d}(x)$}
\label{singbehavchi3}

Now that we have obtained all the solutions of the linear 
differential
equations of $\, {\tilde \chi}^{(3)}_{d}(x)$ and 
$\, {\tilde \chi}^{(4)}_{d}(t)$, analytic at the origin,
 we turn to the exact computation
of their singular behavior at the finite singular points.

To obtain the singular behavior of ${\tilde \chi}^{(3)}_{d}(x)$
 amounts to calculating 
the singular behavior of each term in (\ref{chisum2text}).
The details are given in \ref{Detailed}.

\subsection{The behavior of  ${\tilde \chi}^{(3)}_{d}(x)$ 
as $x  \, \rightarrow \, 1$}
\vskip .1cm 

The evaluation of the singular behavior as $x  \, \rightarrow \, 1$ 
corresponds to straightforward calculations that are given by (\ref{chi1})
(see (\ref{chi2x1}) and (\ref{chi3x1})): 
\begin{eqnarray}
\hspace{-0.8in} Sol( L^{(3)}_6) (Singular, x=1) \, \,\,=\, \, \, \, \, \, 
 {2 \over \pi} \cdot {\frac{3\,a_3^{(3)}\,+a_2^{(3)}}{(1\,-x)^2}}\,\, \,
\,    + \Bigl(a_1^{(3)}\,\,  \,  - \, ({\frac{3a_3^{(3)}\,+a_2^{(3)}}{\pi}}) 
 \nonumber \\
\hspace{-0.7in}  + 3 \,a_2^{(3)} \cdot \, ( \frac{5\pi}{9\Gamma^2(5/6)\Gamma^2(2/3)}
\,-\frac{8\pi}{\Gamma^2(1/6)\Gamma^2(1/3)} ) \Bigr)\cdot \,  {1 \over {1-x}}
 \,  \, \,+{a_2^{(3)} \over {2\pi}} \cdot \ln(1-x).
\end{eqnarray}
When specialized to the combination (\ref{chisum2text})
 defining ${\tilde \chi}^{(3)}_{d}(x)$, the singular
behavior reads
\begin{eqnarray}
\hspace{-0.8in}  {\tilde \chi}^{(3)}_{d}(x)(Singular, x=1)
 \,\,= \,\, \,\Bigl( {1 \over 3} \, 
- \frac{5\pi}{18\Gamma^2(5/6)\Gamma^2(2/3)}\, 
+\frac{4\pi}{\Gamma^2(1/6)\Gamma^2(1/3)}  \Bigr) \cdot {1 \over {1\, -x}}
 \nonumber \\
\hspace{-0.4in}  \qquad \qquad  \qquad  +{1 \over {4\, \pi}} \cdot \ln(1\, -x).
\end{eqnarray}
This result agrees with the result determined numerically
 in Appendix B of~\cite{mccoy3}.

One remarks, for the particular combination (\ref{chisum2text})
 giving $\, {\tilde \chi}^{(3)}_{d}(x)$,  that the {\em most divergent
term disappears}. Note that this is what has been obtained~\cite{jm5}
for the susceptibility  ${\tilde \chi}^{(3)}$ where the
 singularity $(1-4 w)^{-3/2}$ of the ODE is not 
present in ${\tilde \chi}^{(3)}$.

\subsection{The behavior of  ${\tilde \chi}^{(3)}_{d}(x)$ as $x  \, \rightarrow \, -1$}
\label{Thecalcchi3}
\vskip .1cm 

The calculations of the singular behavior as $x  \, \rightarrow \, -1$ 
rely mostly on connection formulae of $\, _2F_1$ hypergeometric functions,
and the results are given below in (\ref{chi2final}) and (\ref{chi3final}).
For the combination (\ref{chisum2text}), the singular behavior reads
\begin{eqnarray}
\hspace{-0.9in} {\tilde \chi}^{(3)}_{d}(x)(Singular, x=-1)
 \,\,=  \, \,\, \,  {1 \over 4\pi^2} \cdot \ln(1+x)^2 \,  \,
+\left( {1 \over 4\pi}\,  -{\frac{2\ln(2)-1}{2\, \pi^2}} \right) \cdot  \ln(1+x),
 \nonumber 
\end{eqnarray}
which agrees with the result of Appendix B of~\cite{mccoy3}.

\subsection{The behavior of  ${\tilde \chi}^{(3)}_{d}(x)$ as 
$x \, \rightarrow \, e^{\pm 2 \pi i/3}$}
\vskip .1cm 
The result for the singular behavior  $\, {\tilde \chi}^{(3)}_{d}(x)$
 as  $\, x \, \rightarrow \,  x_0=e^{\pm 2 \pi i/3}$ reads:
\begin{eqnarray}
&&\hspace{-0.5in}{\tilde \chi}^{(3)}_d (Singular, x=x_0)
\,\,\, = \,\, \, \, \,-\frac{8 \cdot 3^{1/4}}{35\pi} \,e^{\pi i/12} \cdot \,(x-x_0)^{7/2}
 \nonumber \\
&&\hspace{-0.5in} \qquad \quad \, = \,\,\,\, 
-0.0957529 \,\, \cdots \,\, \,e^{\pi i/12} \cdot \,(x-x_0)^{7/2}.
\end{eqnarray}

This result is in agreement with the numerical result of Appendix B of~\cite{mccoy3}
namely $\,-\,\sqrt{2}/3\,\, e^{\pi i/12} \cdot \,b \cdot (x-x_0)^{7/2}$,
with  $ \, \, b \, = \,\,  0.203122784 \, \cdots$

\section{Singular behavior of  ${\tilde \chi}^{(4)}_{d}(x)$}
\label{singbehavchi4}

To obtain the singular behavior of ${\tilde \chi}^{(4)}_{d}(x)$ amounts to obtaining 
the singular behavior of each term in (\ref{chisumL8}). 

\subsection{Behavior of $\,{\tilde \chi}_{d}^{(4)}(t)$  as $t \, \rightarrow \, 1$}
\label{behavi}
\label{Thecalcchi4}

The calculations of the singular behavior of $\,{\tilde \chi}^{(4)}_{d;2}(t)$ as 
$t\, \rightarrow\, 1$ are displayed in \ref{detailchi4}, and read
\begin{eqnarray}
\label{4chi2final}
&&{\tilde \chi}^{(4)}_{d;2}(t)(Singular, t=1)
 \,\,\,\,=\,\,\,\,\,
\frac{8}{\pi^2(1-t)^2}\,\,\,\,\,-\frac{8}{\pi^2(1-t)}\,\, 
\nonumber \\
&& \qquad  \qquad  \qquad +\frac{5}{2\pi^2} \cdot \ln\frac{16}{1-t}\,\,\,\,
-\frac{3}{2\pi^2}\cdot \ln^2\frac{16}{1-t}.
\end{eqnarray}

To compute the singular behavior of $\,{\tilde \chi}^{(4)}_{d;3}(t)$
 as $t\,\rightarrow\, 1$ we
need the expression of the hypergeometric function
${}_4F_3([1/2,\,1/2,\,1/2,\,1/2],[1,\,1,\,1];z)$ as
$z\rightarrow 1$. This hypergeometric function is an example of solution of a 
Calabi-Yau ODE, and explicit computations of its monodromy
matrices have been given~\cite{chen}.

The differential equation for
 ${}_4F_3([1/2,\,1/2,\,1/2,\,1/2],[1,\,1,\,1];z)$ 
is {\em Saalsch{\"u}tzian and well-poised}
 (but not very-well-poised). At
$z\,=\,1$ it has one logarithmic solution and
three analytic solutions of the form 
\begin{equation}
\sum_{n=0}^{\infty}\, c_n \cdot (1-z)^n.
\end{equation}
The $c_n$ satisfy the fourth order recursion relation
\begin{eqnarray}
&&\hspace{-0.5in}16\,n \cdot (n-1)^2 \, (n-2) \cdot c_{n}
 \,\,\, -24\,(n-1)(n-2)(2n^2-6n+5) \cdot c_{n-1}
\nonumber\\
&&\hspace{-0.5in}\quad +16\,(n-2)^2\,(3n^2-12n +13) \cdot c_{n-2} \,
 \, \, -(2n-5)^4 \cdot c_{n-3} \,\,\,\, = \,\,\,\, \, 0, 
\end{eqnarray}
where $c_n= \, 0$ for $n \leq \, -1$. The vanishing of the coefficient 
$c_n$ at $n= \, 0, \, 1, \, 2$, of $c_{n-1}$ at $n = \, 1,\, 2$
 and $c_{n-2}$ at $n= \, 2$
guarantees that $c_0,~c_1,~c_2$ may be chosen arbitrarily.

The behavior at $z= \, 1$ of   ${}_4F_3([1/2,1/2,1/2,1/2], [1,1,1];z)$, which is the
solution of the ODE that is analytic at $z\,=\, 0$, 
is given in Theorem 3 of B{\"u}hring~\cite{buhring} 
with the parameter 
\begin{equation}
\label{sdef}
s \, \,= \, \, \,\, \,\sum_{j=1}^3 \,b_j \, \,\, -\sum_{j=1}^4 \,a_j \,\, \,= \, \,\,1, 
\end{equation}
(i.e. the {\em Saalsch{\"u}tzian condition}~\cite{Saal,Saal2,Saal3}).
For completeness we quote this theorem which is valid for all
${}_{p+1}F_p([a_1, \, \cdots, \, a_{p+1}], \, [b_1, \, \cdots, \, b_p]; \, z)$
 when the parameter
$s$ of (\ref{sdef}) is {\em any integer}\footnote[2]{Again
 we emphasise the role of 
$\, k$-balanced hypergeometric functions.} $s\geq 0$:
\begin{eqnarray}
\label{buhring}
&&\hspace{-0.8in}\frac{\Gamma(a_1)\cdots
\Gamma(a_{p+1})}{\Gamma(b_1)\cdots\Gamma(b_p)}
\cdot \, 
{}_{p+1}F_p([a_1,\cdots,a_{p+1}], \,[b_1,\cdots,b_p]; \, z)
\nonumber \\
&& \,=\,\,\,\,\,\,
\sum_{n=0}^{s-1}\,I^<_n\cdot (1-z)^n\,\,
+\sum_{n=s}^{\infty}\, I^>_n \cdot (1-z)^n 
\nonumber  \\
&&\quad \quad \,\,+(1-z)^s \cdot \sum_{n=0}^{\infty}
[w_n\,\, +q_n \cdot \ln(1-z)]\cdot (1-z)^n, 
\end{eqnarray}
for $|1-z|\,< \, 1, \,-\pi\,<\,arg(1-z)\,<\,\pi$ and
 $\,p\,= \, 2,\,3,\, \cdots$
where for $\,0\,\leq\, n\,\leq\, s-1$
\begin{equation}
\hspace{-0.05in}I^<_n \,\, = \, \,\, \, 
 (-1)^n \cdot \frac{\Gamma(a_1+n)\Gamma(a_2+n)(s-n-1)!}
{\Gamma(a_1+s)\Gamma(a_2+s)n!} \cdot \sum_{k=0}^{\infty}
\frac{(s-n)_k}{(a_1+s)_k\, (a_2+s)_k} \cdot \, A^{(p)}_k, 
\end{equation}
for $s\leq n$
\begin{equation}
\hspace{-0.05in}I^>_n \,\,\,  = \,\, \,\, 
 (-1)^n \cdot \frac{(a_1+s)_{n-s}\, (a_2+s)_{n-s}}{n!} \cdot 
\sum_{k=n-s+1}^{\infty}\, 
\frac{(k-n+s)!}{(a_1+s)_k\,(a_2+s)_k} \cdot A^{(p)}_k, 
\end{equation}
and $\,w_n$ and $\,q_n$ are such that
\begin{eqnarray}
\label{qn}
&&\hspace{-0.8in} w_n\,+q_n \cdot \ln(1-z)
\,\,\,=\,\,\,\,\, (-1)^s \cdot \frac{(a_1+s)_n\,(a_2+s)_n}{(s+n)!\, n!}
\nonumber\\
&  \times& \Bigl(\sum_{k=0}^n\frac{(-n)_k}{(a_1+s)_k(a_2+s)_k} \cdot A^{(p)}_k
\cdot \, \Big[\psi(1+n-k) +\psi(1+s+n) 
\nonumber\\
&& \quad \quad \quad  -\psi(a_1+s+n)\, -\psi(a_2+s+n)\, -\ln(1-z)\Big]
\Bigr),
\end{eqnarray}
where $(a)_n\,=\, \, a \, (a+1)\, \cdots \,(a+n-1)$ 
is the Pochhammer's symbol.
The $A^{(p)}_k$ are computed recursively in~\cite{buhring} as $p-1$
fold sums. In particular
\begin{equation}
A_k^{(2)}\,\,=\,\, \,\,  \,\frac{(b_2-a_3)_k\, (b_1-a_3)_k}{k!}, 
\end{equation}
and
\begin{eqnarray}
\label{a3k}
&&\hspace{-1in}A^{(3)}_k\,\,=\,\,\, \,
\sum_{k_2=0}^k\, \frac{(b_3+b_2-a_4-a_3+k_2)_{k-k_2}\, (b_1-a_3)_{k-k_2}\,
(b_3-a_4)_{k_2}\,(b_2-a_4)_{k_2}}{(k-k_2)! \, k_2!}
\nonumber\\
&&\hspace{-0.8in} \quad  \quad = \,\,\, \, 
\frac{(b_1+b_3-a_3-a_4)_k \,(b_2+b_3-a_3-a_4)_k}{k!}
\\
&& \,  \times \, 
{}_3F_2([b_3-a_3, \, b_3-a_4, \, -k],\, [b_1+b_3-a_3 -a_4, \, b_2+b_3-a_3-a_4], \, 1).
\nonumber
\end{eqnarray}

For use in (\ref{chi43}) we need to 
specialize to  $a_j =\, 1/2,~b_j= \, 1$,  where
\begin{eqnarray}
&&A^{(3)}_k\, \,  =\,\,\, \,    \sum_{k_2=0}^k \, 
\frac{(1+k_2)_{k-k_2}\, (1/2)_{k-k_2}\,  (1/2)_{k_2}^2}{(k-k_2)! \, k_2!}
 \nonumber \\
&& \qquad \quad \quad \,= \, \,\, \,  \, \, 
  k! \cdot \, {}_3F_2([1/2,1/2,-k],\, [1,1]; \, 1), 
\end{eqnarray}
and for respectively $\, n=0$ and  $n\geq 1$
\begin{equation}
\label{indef}
\hspace{-0.05in} I^<_0 \,\, = \, \, \, 
4 \, \sum_{k=0}^{\infty} \, \frac{k!}{(3/2)^2_k} \cdot \, A^{(3)}_k,
\quad \quad 
I^>_n\,\,  =\,\,\,   \, (-1)^{n} \,\frac{(3/2)_{n-1}^2}{(n)!}
\cdot \sum_{k=n}^{\infty}\frac{(k-n)!}{(3/2)_k^2}\cdot A^{(3)}_k. 
\end{equation}
We note, in particular, the terms
\begin{eqnarray}
A^{(3)}_0\,=\,1,\,\qquad A^{(3)}_1\,\,=\,\,3/4, \qquad A^{(3)}_2\,=\,\,41/32.
\end{eqnarray}
Using these specializations in (\ref{buhring}) we compute the terms in
$\,{\tilde \chi}^{(4)}_{d;3}(t)$ which diverge as $\,t\,\rightarrow\, 1$. The
term $ \, (1-t)^{-1} \cdot \ln(1-t^2)$ cancels and  we are left with
\begin{eqnarray}
\label{4chi3final}
\hspace{-0.99in} {\tilde \chi}^{(4)}_{d;3}(t) (Singular, t=1)\,  =\,\,
\frac{1}{\pi^2}\cdot \left(\frac{8}{3 \, (1-t)^2}
\,\,+\frac{56}{3\, (1-t)}\,+\frac{16}{3 \cdot (1-t)} 
\cdot (3\,I_1^{>} \,-4\,I_2^{>})\right) 
\nonumber \\
\hspace{-0.8in} \qquad  \qquad \qquad \qquad \qquad \qquad \,\,
+\frac{8}{3 \, \pi^2} \cdot  \ln\frac{1-t^2}{16}.
\end{eqnarray}

Thus, using (\ref{4chi1}), (\ref{4chi2final}) and (\ref{4chi3final}),  
we find that the terms in $\, sol ( L^{(4)}_8 )$, which diverge as
 $\, t \, \rightarrow\,  1$, are
\begin{eqnarray}
\hspace{-0.8in} Sol ( L^{(4)}_8 ) (Singular, t=1) \,\,\,\, =\,\,\,\,
 {\frac{8 \, (a_3^{(4)}+3a_2^{(4)})}{8 \, \pi^2}} \cdot  {\frac{1}{(1-t)^2}} \,
 \nonumber \\
\hspace{-0.8in} \qquad \qquad \quad
 + \Bigl(a_1^{(4)} \, -{\frac{8(3a_2^{(4)}\, -7a_3^{(4)})}{3\pi^2}}\,\, 
+{\frac{16\,a_3^{(4)}}{3\pi^2}}\cdot  (3I^>_1\,-\,4I^>_2)   \Bigr)\cdot   {\frac{1}{1-t}} 
\nonumber \\
\hspace{-0.8in} \qquad \qquad \quad
+\, {\frac{15a_2^{(4)}-16a_3^{(4)}}{\pi^2}}\cdot \ln ({\frac{16}{1-t}}) \, \,  \, \, 
-{\frac{3a_2^{(4)}}{2\pi^2}}\cdot \ln^2 ({\frac{16}{1-t}}),
\end{eqnarray}
where the constant $\, 3I^>_1\,-\, 4I^>_2$ reads (with 200 digits):
\begin{eqnarray}
\label{guess}
\fl \qquad \qquad 3I^>_1\,-\, 4I^>_2 \,\,\,\,\, =\,\, 
\nonumber \\
\fl \qquad \qquad 
-2.212812128930821923547976814986050021481359293357467766171  
\nonumber \\
\fl \qquad \qquad 
630847360232164854964985815375185842526324049358792616932061 
\nonumber \\
\fl \qquad \qquad 
331297671076950376704358248264961101007730925578212714241825 
\nonumber \\
\fl \qquad \qquad 
5205323181711923135264 \, \cdots
\end{eqnarray}

When specializing to the particular combination (\ref{chi4exp}), the singular 
behavior of the integral ${\tilde \chi}^{(4)}_d(t)$ reads
\begin{eqnarray}
\fl \qquad {\tilde \chi}^{(4)}_d(t)(Singular, t=\,1)\, \,\,=\,\,\,
\frac{1}{8\,(1-t)}
 \cdot \left(1\, -\frac{1}{3\pi^2}[64\,+16 \cdot (3I^>_1\,-\,4I^>_2)]\right)\,
\nonumber\\
\fl \qquad \qquad \qquad\qquad\qquad \quad \quad 
 +\frac{7}{16\pi^2}\cdot \ln\frac{16}{1-t}\,\,\,\,\,\,
\,-\frac{1}{16\pi^2}\cdot \ln^2\frac{16}{1-t}, 
\end{eqnarray}
This agrees\footnote{Note that there is
 an overall factor of 2 between this result and the
results given in Appendix B of~\cite{mccoy3}
 which comes from a multiplicative factor
of 2 in the series (around $t=\, 0$) of
 ${\tilde \chi}^{(4)}_d(t)$ used in~\cite{mccoy3}.
This applies also to the result of the singular behavior at $t=\, -1$.}
with the result determined numerically in Appendix B of~\cite{mccoy3}.

We find again and similarly to ${\tilde \chi}^{(3)}_d(t)$ that
 the most divergent term disappears
for the particular combination giving ${\tilde \chi}^{(4)}_d(t)$. 
And here again, this is what has been observed~\cite{jm5} for the 
susceptibility ${\tilde \chi}^{(4)}$ at the singularity $x=\,16w^2=\, 1$
 which occurs in the ODE as $z^{-3/2}$
and cancels in the integral ${\tilde \chi}^{(4)}$.

\vskip .3cm
{\bf Remark:}  It is worth recalling that similar calculations for
$\, {\tilde \chi}^{(4)}$, also based on the evaluation of a 
connection matrix (see section 9 of~\cite{jm4}), 
require the evaluation of a constant $\, I_4^{-}$ that is actually expressed 
in terms of $\zeta(3)$:
\begin{equation}
\label{I4minus}
I_4^{-} \,\,\, = \,\,\,\, 
{{1} \over {16 \, \pi^3 }} \cdot \Bigl({{4 \, \pi^2 } \over { 9}}
 \,  -  {{1} \over { 6}}  \, -\, {{7} \over {2}} \cdot \zeta(3)\Bigr),  
\end{equation}
when the bulk $\, {\tilde \chi}^{(3)}$ requires 
a Clausen constant~\cite{jm4} that can be written as: 
\begin{equation}
\label{Clau}
Cl_2(\pi/3) \, \,\, = \, \,\, \,\, {{3^{1/2}} \over {108}} \cdot 
(3 \cdot \psi(1, \, 1/3) \, + 3 \cdot \psi(1, \, 1/6) \, -8 \, \pi^2).
\end{equation}
It is quite natural to see if the constant $\, 3I^>_1\,-\,4I^>_2$ 
given with 200 digits in (\ref{guess}), can also be obtained exactly 
in terms of known transcendental constants ($\, \zeta(3), \, \cdots$),
 or evaluations of hypergeometric functions that naturally occur in 
connection matrices~\cite{jm4} (see (\ref{4F3t2eval}) in \ref{zeta}).
This question is sketched in \ref{zeta}.

\subsection{Behavior of $\,{\tilde \chi}_{d}^{(4)}(t)$
  as $t \, \rightarrow \, -1$}
\label{behavtminusone}

When $t\,\rightarrow \,-1$ the only singular terms come from
${\tilde \chi}^{(4)}_{d;3}(t)$. Furthermore 
the operator $\,A_3$ of (\ref{A3}) is non-singular at $t=-1$. Therefore,
 the only singularities in $\,{\tilde \chi}^{(4)}_d(t)$ come from the terms with
$\ln(1-t^2)$  in the expansion (\ref{buhring}) of 
${}_4F_3([1/2,1/2,1/2,1/2],[1,1,1]; \, t^2)$ at
 $\,t\,\rightarrow\, -1$. Thus, from
(\ref{chi4exp}) we find that the singular part of 
$ \, {\tilde \chi}^{(4)}_d(t)$ at $\, t \,=\,-1$ reads
\begin{equation}
{\tilde \chi}^{(4)}_{d;sing}(t) \,= \,  \,\,
 -\frac{1}{8}\, {\tilde \chi}^{(4)}_{d;3;sing}(t)\,
 \, \,= \, \, \, 
-\frac{1}{8}\ln(1-t^2) \cdot
\, A_3 \Bigl( \sum_{n=0}^{\infty} \,q_n \cdot (1-t^2)^{n+1}\Bigr), 
\end{equation}
with $ \, q_n$ obtained from (\ref{qn}) as
\begin{equation}
q_n \,\,= \, \,\, \, \,
\frac{(3/2)^2_n}{(n+1)!\, n!} \cdot 
\sum_{k=0}^n \,\frac{(-n)_k}{(3/2)^2_k} \cdot \, A^{(3)}_k,
\end{equation}
where $\,  A^{(3)}_k$ is given by (\ref{a3k}).
We know  from the exponents of $L^{(4)}_4$ at $t\,  =\,  -1$ 
that the result has the form
$\,(t+1)^7 \cdot \ln(t+1)$. Therefore to obtain this term
 in a straightforward
way we need to expand the coefficient of $\ln(1-t^2)$ to order $ \,(1+t)^{9}$
in order that the term from $\,(1+t)\cdot D_t^3$ be of order $\,(1+t)^7$. This is
tedious  by hand but is easily done on Maple and we find that
the leading singularity in $ \,{\tilde \chi}^{(4)}_d(t)$ at $t=\, -1$ is
\begin{eqnarray}
{\tilde \chi}^{(4)}_{d}(Singular, t=-1)
 \, \, \,= \, \, \,\,\frac{1}{26880}\cdot (1+t)^7 \cdot \, \ln(1+t), 
\end{eqnarray}
which agrees with Appendix B of~\cite{mccoy3}.

\section{Conclusion: is the Ising model ``modularity''
reducible to selected $\, _{(q+1)}F_q$ hypergeometric functions ?}
\label{concl}

In this paper we have derived the exact analytic expressions for
${\tilde \chi}^{(3)}_d(x)$ and
 ${\tilde \chi}^{(4)}_d(t)$ and from them have computed
the behavior at all singular points. We have also 
obtained some additional exact results for  
${\tilde \chi}^{(5)}_d(x)$ (see section \ref{chi5}).
This completes the program
initiated in~\cite{mccoy3} where the singularities were studied by
means of formal solutions found on Maple and numerical studies of the
connection problem~\cite{jm4}. In this sense we have a complete solution to the
problem. However, in another sense, there are still most interesting open
questions. 

In section \ref{Thecalcchi4} we used the solution
 of the hypergeometric connection
problem ~\cite{buhring} which gave the connection constants $\,I^<_n$ and 
$\,I^>_n$ as multiple sums. However there are  special cases, as
mentioned in~\cite{buhring2}, where it is known by indirect means that 
the series can be simplified, but for which a direct simplification
of the series has not been found. One example is given by the
computation in section \ref{Thecalcchi3} of the singularity
 of $\,  {\tilde \chi}^{(3)}_d(x)$ at $\,  t\,  =\,1$
which we accomplished by means of the reduction (\ref{Prud}) of a
${}_3F_2$ function to a product of ${}_2F_1$ functions. This produced
the gamma function evaluation  of the singularity at $x\,  =\,  1$ of 
(\ref{chi3x1}). This singularity could also have been computed
directly from the ${}_3F_2$ function in (\ref{chi31}) by use of the
B{\"u}hring formula (\ref{buhring}) but a reduction of the sums for
the required $I_n$ to the gamma function form is lacking. There are
 two suggestions that such a reduction may exist for
$\,{\tilde \chi}^{(4)}_d(t)$ at $\,t\,= \, 1$. The first is that, by analogy 
with the corresponding calculation for ${\tilde \chi}^{(4)}(t)$ 
in the bulk~\cite{tracy}, the amplitude
could be evaluated in terms of $\,\zeta(3)$. The second is that
evaluations of Calabi-Yau~\cite{chen} hypergeometric functions like
 ${}_4F_3([1/2,\,1/2,\,1/2,\,1/2],[1,\,1,\,1],\,\,z)$  take place. 
The larger question, of course, is how much the structure seen in
$\,{\tilde \chi}^{(n)}_d$ and $\,{\tilde \chi}^{(n)}$ for 
$n\,=\,1,\,2,\,3,\,4$ can be expected
to generalize to higher values of $n$. 
It is the opinion of the authors that there is a great deal of mathematical
structure of deep significance remaining to be discovered.   
 
These new exact results for the diagonal susceptibility 
of the Ising model confirm that the linear differential operators
that emerge in the study of these Ising $\, n$-fold integrals,
are {\em not only} ``Derived From Geometry''~\cite{jm6}, but actually 
correspond to ``Special Geometries'' (they are homomorphic to their adjoints, 
which means~\cite{Katz} that their differential Galois groups
 are ``special'', their 
symmetric square, exterior square has rational function solutions, ...). 
More specifically, when we are able to get 
the exact expressions of these linear
 differential operators, we find out that they
 are associated 
with elliptic function theory ({\em elliptic functions}~\cite{amm}
{\em or modular forms}),
and, in more complicated cases, 
{\em Calabi-Yau} ODEs~\cite{Almkvist,TablesCalabi}. This totally confirms
what we already saw~\cite{jm6} on $\, {\tilde \chi}^{(5)}$ 
and  $\, {\tilde \chi}^{(6)}$. 
 We see in particular, with $\, \chi^{(5)}_d(x)$, the emergence
of a remarkable order-six operator which is such that
{\em its symmetric square has a rational solution}. 

Let us recall that it is, generically, extremely difficult to see 
that a linear differential operator corresponding to 
 a Calabi-Yau ODE~\cite{Almkvist,TablesCalabi},
 is homomorphic to a $\, _{q+1}F_q$ 
hypergeometric linear differential operator up to an algebraic pullback.
Worse, it is not impossible that many of the Calabi-Yau ODEs are 
actually reducible (up to operator equivalence) to $\, _{q+1}F_q$ 
hypergeometric functions up to  algebraic pullbacks
 that have not been found yet.
Let us assume that this is not the case, and that the Calabi-Yau world 
is not reducible to the hypergeometric world (up to involved
algebraic pullback), we still have to see if the ``Special Geometry''
 operators that occur for the Ising model, are ``hypergeometric'' ones,
reducing, in fact systematically to (selected $k$-balanced)
$\, _{q+1}F_q$ hypergeometric functions, or correspond to the more
general solutions of Calabi-Yau equations.

\vspace{.1in}

\vskip .5cm 

{\bf Acknowledgment}
This work was supported in part by the National Science Foundation
grant PHY-0969739. We thank  D. Bertrand, A. Bostan, J. Morgan, A. Okunkov, and  
J-A. Weil for fruitful discussions. M. van Hoeij is 
supported by NSF grant 1017880.
 This work has been performed without any support
 of the ANR, the ERC or the MAE. 

\appendix

\section{Miscellaneous comments on the modular curve (\ref{modularcurv})}
\label{miscell}

Let us introduce other rational expressions, similar to (\ref{qdef}) 
and (\ref{qdef2}):
\begin{eqnarray}
\label{qdef2bis}
&&Q_2(x) \,  = \,  \,{{27\, x^4 \cdot  (1+x) } \over { (x\, +\, 2)^6}}, 
\qquad 
\, \,\,\,\,\,
Q_3(x) \, = \, \,  -\, {{27\, x \cdot (1+x)^4 } \over {(x-1)^6 }}, 
\nonumber 
\end{eqnarray}
where recalling the expression of (\ref{qdef2}) one has (for instance): 
\begin{eqnarray}
&&\hspace{-0.5in}Q_2(x) \, \, = \, \, \,Q_1\Bigl( {{1} \over {x}} \Bigr) 
 \, \, = \, \, \,Q_1\Bigl( -\, {{1\, +x} \over {x}} \Bigr), \, 
\qquad \quad Q_3(x) \,\, \, = \, \, \,Q_1\Bigl( -\, {{1} \over {1\, +\, x}} \Bigr) 
\nonumber \\
&&  \, \, = \, \, \,Q_1\Bigl( -\, {{x} \over {1\, +\, x}} \Bigr)
\,\, \, = \, \, \, Q_2(-1\, -x) 
 \,\, \, = \, \, \, Q_2\Bigl( -\, {{1\, +x} \over {x}} \Bigr).
\nonumber
\end{eqnarray}

Remarkably the elimination of $\, x$ between 
the Hauptmodul $\, Q \, = \, \, Q(x)$ and
$\, Q_2 \, = \, \, Q_2(x)$
(or  $\, Q \, = \, \, Q(x)$ and
$\, Q_3 \, = \, \, Q_3(x)$) also gives the {\em same} modular curve 
(\ref{modularcurv}).

We also have remarkable identity on 
the {\em same} hypergeometric function 
 with these new Hauptmodul pullbacks (\ref{qdef2bis}):
\begin{eqnarray}
\label{bingo2}
&& (x\, +2)
\cdot 
\, _2F_1\Bigl([{{1} \over {6}}, \,{{1} \over {3}}], \, [1]; \,Q(x)\Bigr)
\nonumber \\
&& \qquad \quad \, \, = \, \, \, \, 2 \cdot (1+x+x^2)^{1/2}
\cdot 
 \, _2F_1\Bigl([{{1} \over {6}}, \,{{1} \over {3}}], \, [1]; \,Q_2(x)\Bigr), 
\end{eqnarray}
and:
\begin{eqnarray}
\label{bingo3}
&& (1\, -x)
\cdot 
\, _2F_1\Bigl([{{1} \over {6}}, \,{{1} \over {3}}], \, [1]; \,Q(x)\Bigr)
\nonumber \\
&& \qquad \quad \, \, = \, \, \, \, (1+x+x^2)^{1/2}
\cdot 
 \, _2F_1\Bigl([{{1} \over {6}}, \,{{1} \over {3}}], \, [1]; \,Q_3(x)\Bigr). 
\end{eqnarray}

The well-known fundamental modular 
curve~\cite{jm9}
\begin{eqnarray} 
\label{fundmodular}
&&5^9\, v^3\, u^3\, -12 \cdot 5^6 \, u^2\, v^2 \cdot (u+v)\, 
+375\, \, u\, v \cdot  (16\, u^2\, +16\, v^2\, -4027\, v\, u)
 \nonumber \\
&&\quad \quad -64\, (v+u)\cdot (v^2+1487\, v\, u\, +u^2) \, \, 
+2^{12}\cdot  3^3 \cdot  u\, v  \,\,  = \, \, \, 0,
\end{eqnarray}
 corresponding to the elimination
 of the variable  $\, x$ between the previous 
Hauptmodul (\ref{qdef}) and another 
Hauptmodul\footnote[3]{Related by a Landen
 transformation on  $\, x^{1/2}$ see~\cite{jm9}.} $\, Q_L(x)$:
\begin{eqnarray} 
\label{QL}
Q_L(x) \,  \, = \,  \, \, 
-108 \cdot \frac{(1\, +x)^4 \cdot x}{(x^2\, -14 \, x\, +1)^3}.
\end{eqnarray}
should not be confused with the 
 (modular) curve~\cite{jm9} (\ref{modularcurv}).

The new modular curve (\ref{modularcurv})
also has a rational parametrization, 
 $\, (u, \, v) \, = \, \, (Q_L(x), \, Q_4(x))$, 
between this last new Hauptmodul (\ref{QL}) and a new 
simple  Hauptmodul:
\begin{eqnarray} 
Q_4(x) \,  \, = \,  \, \, 
108 \cdot {{(1\, +x)^2 \cdot x^2 } \over {(1\, -x)^6 }}.
\end{eqnarray}

\section{Solution of  $\, {\mathcal M}_4$ analytical at $\, x\, = \, \, 0$}
\label{M4}

The solution of  $ \, {\mathcal M}_4$  (see (\ref{l44other})),
 analytical at $\, x\, = \, \, 0$, reads:
\begin{eqnarray}
\label{solu}
\hspace{-0.8in} Sol(x) \, \,  = \, \,\,   (1\, -x)^{3/4} \cdot \rho(x) \cdot S(x), 
\end{eqnarray}
where $\, S(x)$ reads: 
\begin{eqnarray}
&&\hspace{-0.9in} Z_1 \cdot \, 
 _4F_3\Bigl([{{1} \over {2}}, \, {{1} \over {2}}, \, 
{{1} \over {2}}, \, {{1} \over {2}}], \, [1, \, 1, \, 1]; \, P(x)\Bigr) \, \, 
 +Z_2 \cdot \, 
 _4F_3\Bigl([{{3} \over {2}}, \, {{3} \over {2}}, \, {{3} \over {2}}, \, {{3} \over {2}}], 
\, [2, \, 2, \, 2]; \, P(x)\Bigr)
\nonumber \\
&&\hspace{-0.9in}  +Z_3 \cdot \, 
 _4F_3\Bigl([{{5} \over {2}}, \, {{5} \over {2}}, \, {{5} \over {2}}, \, {{5} \over {2}}], 
\, [3, \, 3, \, 3]; \, P(x)\Bigr) \, \, 
+Z_4 \cdot \, 
 _4F_3\Bigl([{{7} \over {2}}, \, {{7} \over {2}}, \, {{7} \over {2}}, \, {{7} \over {2}}], 
\, [4, \, 4, \, 4]; \, P(x)\Bigr),
\nonumber 
\end{eqnarray}
with 
\begin{eqnarray}
\hspace{-0.8in} Z_1 \, \, = \, \, \, -512 \cdot {{n_1} \over {d_1}}, \quad \, 
Z_2 \, \, = \, \, \, 128 \cdot {{n_2} \over {d_2}}, \quad \, 
  Z_3 \, \, = \, \, \, -54 \cdot {{n_3} \over {d_3}}, \quad \, 
 Z_4 \, \, = \, \, \, -625 \cdot {{n_4} \over { d_4}}, \nonumber 
\end{eqnarray}
and
\begin{eqnarray}
&&\hspace{-0.8in}n_1 \, \, = \, \, \,
 (7 \, x^3 \, -56 \, x^2\,+112 \, x\,-64) \,  \cdot \, (1\, -x)^{1/2} 
\nonumber \\
&&\hspace{-0.6in} \qquad 
\,  +(x-1) \, (x^3\,-24 \, x^2\,+80 \, x\,-64),
\nonumber \\
&&\hspace{-0.8in}n_2 \, \, = \, \, \, 
(2352 \, x^2 \,-472 \, x^3 \,-3904 \, x \,+2048 \,+19 \, x^4) \,  \cdot \, (1\, -x)^{1/2}
\nonumber \\
&&\hspace{-0.7in} \qquad  
+x^5 \,-125 \, x^4 \,\,+1288 \, x^3 -4048 \, x^2 \,+4928 \, x  \,-2048,
\nonumber \\
&&\hspace{-0.8in}n_3 \, \, = \, \, \,
 (x^6 \, -28080 \, x^3\,-355 \, x^5\,-52992 \, x\,
+17920+5750 \, x^4\,+57760 \, x^2)  \cdot \, (1\, -x)^{1/2}
 \nonumber \\
&&\hspace{-0.6in} \qquad 
   -2 \, (x-1) \, (20 \, x^5\,-855 \, x^4\,+6736 \, x^3\,
-18992 \, x^2\,+22016 \, x\,-8960), 
\nonumber \\
&&\hspace{-0.8in}n_4 \, \, = \, \, \,(x^2\,-8 \, x\,+8) \, 
(x^4\, -64 \, x^3\,+320 \, x^2\,-512 \, x\,+256) \cdot \, (1\, -x)^{1/2}
\nonumber \\
&&\hspace{-0.6in} \qquad  
     -4 \, (x-1) \, (x-2) \, (3 \, x\,-4) \, (x\,-4) \, (x^2\,-16 \, x\,+16),
 \nonumber 
\end{eqnarray}
and
\begin{eqnarray}
&& d_1 \,\,  \, = \, \, \,(1-x)\cdot x^2 \cdot
 ((x-2)\cdot (x^2\,-16\,x\,+16) \cdot (1\, -x)^{1/2}\,
\nonumber \\
&& \qquad \quad -2\,(x-1)\,(3\,x-4)\,(x-4) ),
  \nonumber \\
&& d_2 \,\,  \, = \, \, \,(1-x)\cdot x^4 \cdot 
(4\cdot (x-2) \cdot (1\, -x)^{1/2} \, + \, x^2\,-8\, x\,+8),
\nonumber \\
&&d_3 \,\,  \, = \, \, \,(1-x)\cdot x^6 \cdot (2\, (x-1) \, -(x-2) \cdot (1\, -x)^{1/2}   ),
 \nonumber \\
&&d_4 \,\,  \, = \, \, \, (1-x)\cdot x^8, 
\nonumber 
\end{eqnarray}
and 
\begin{eqnarray}
\rho(x) \, \, = \, \, \, 
\Bigl( (2 - x) \cdot (1\, -x)^{1/2} \,  + 2 \cdot (x\, -1)  \Bigr)^{1/2}, 
 \nonumber
\end{eqnarray}
and where $\, P(x)$ denotes 
the pullback (\ref{Pull}): 
\begin{eqnarray}
\label{Pullbis}
P(x) \, \, = \, \, \,\, \,
 {{x^2 \, -8 \, x \, +8} \over {x^2}} \,\, \, \,
 -4 \cdot (2-x) \cdot {{(1-x)^{1/2} } \over {x^2}}.
\end{eqnarray}

This solution has the integrality property~\cite{Kratten}. Changing 
$\,x$ into $\, 64 \, x$
the series expansion of the previous solution (\ref{solu})
 has {\em integer coefficients}:
\begin{eqnarray}
&&\hspace{-0.4in} Sol(64\, x) \, \,  = \, \,\,  128\, \, +2560 \, x\, +116736 \, x^2\, 
+6072320 \, x^3\, +335104000 \, x^4\,
\nonumber \\
&&\hspace{-0.4in} \quad \quad \quad +19117744128 \, x^5
 +1114027622400 \, x^6+65874638708736 \, x^7 
\nonumber \\
&&\hspace{-0.4in} \quad \quad \quad +3937277209282560 \, x^8  \, + \, \, \cdots 
\nonumber 
\end{eqnarray}

\section{The linear differential operator ${\cal L}^{(5)}_{11}$ in exact arithmetic}
\label{appL11}

The factors occurring in the differential operator ${\cal L}^{(5)}_{11}$ read
\begin{eqnarray}
U^{(5)}_{1} \, \, =\, \, \,   D_x \,\,\, 
- \, {{d} \over {dx}} \ln\Bigl( {{x} \over {(1\, -x)^3}}  \Bigr), 
\end{eqnarray}
\begin{eqnarray}
V^{(5)}_{1} \, = \,\,\,  D_x \,\, \,
- \, {{1} \over {2}} \cdot {{d} \over {dx}} \ln\Bigl( {{(1\, +x \, +x^2)^{3}
} \over {(1+x)^2 \cdot (1\, -x)^6 \cdot x^2}}  \Bigr), 
\end{eqnarray}
\begin{eqnarray}
W^{(5)}_{1} \, = \,\,\, 
D_x \,\,\, - {{1} \over {2}} \cdot 
{{d} \over {dx}} \ln\Bigl(  {{ (x^2+1)^2} \over {(1+x)^2 \cdot (1\, -x)^6 \cdot x }} \Bigr), 
\end{eqnarray}
\begin{eqnarray}
L^{(5)}_{4} \,\, = \, \,\, \,\, D_x^4\,\, \,+{\frac{p_3}{p_4}} \cdot \, D_x^3 \, \, \,
+{\frac{p_2}{p_4}} \cdot \, D_x^2 \, \,+{\frac{p_1}{p_4}} \cdot \,D_x \,
\,\, + {\frac{p_0}{p_4}}, 
\end{eqnarray}
with:
\begin{eqnarray}
\fl p_4 \,=\,   
{x}^{3} \cdot \, (1+x+{x}^{2})  \cdot \, (x+1)^{3} \cdot \, (x-1)^{4} \cdot \,
\Bigl( 160+3148\,x+24988\,{x}^{2}+86008\,{x}^{3}  
\nonumber \\
\fl \quad +141698\,{x}^{4}+69707\,{x}^{5}-141750\,{x}^{6}-358707\,{x}^{7}
-356606\,{x}^{8}-1071\,{x}^{9} +347302\,{x}^{10}
  \nonumber \\
\fl \quad +510214\,{x}^{11}+347302\,{x}^{12}-1071\,{x}^{13}-
356606\,{x}^{14}-358707\,{x}^{15} -141750\,{x}^{16} 
 \nonumber \\
\fl \quad +69707\,{x}^{17}+141698\,{x}^{18}+86008\,{x}^{19}
+24988\,{x}^{20}+3148\,{x}^{21}+160\,{x}^{22} \Bigr),  
 \nonumber 
\end{eqnarray}
\begin{eqnarray}
\fl p_3 \,=\,   
2\,\, {x}^{2}  \cdot \, (x+1)^{2} \cdot \, (x-1)^{3} \cdot
\Bigl( -880-16620\,x-126586\,{x}^{2}-421558\,{x}^{3}-520547\,{x}^{4}
  \nonumber \\
\fl \quad +733378\,{x}^{5}+3794648\,{x}^{6}+6252130\,{x}^{7}
+3922367\,{x}^{8}-4349032\,{x}^{9}
-12817741\,{x}^{10}   
\nonumber \\
\fl \quad -12881692\,{x}^{11}-2612141\,{x}^{12}+10986996
\,{x}^{13}+16830947\,{x}^{14}+12283572\,{x}^{15}  
\nonumber \\
\fl \quad +729267\,{x}^{16}-8919176\,{x}^{17}-10905121\,{x}^{18}
-5398478\,{x}^{19}+866024\,{x}^{20}
  \nonumber \\
\fl \quad +3665682\,{x}^{21}+3069821\,{x}^{22}+1351818\,{x}^{23}+323590\,{x}^{24}
+36308\,{x}^{25}+1680\,{x}^{26} \Bigr),  
 \nonumber 
\end{eqnarray}
\begin{eqnarray}
\fl p_2 \,=\, \,  
2\,\, x  \cdot \left( x-1 \right)^{2} \cdot
\Bigl( 2400+38692\,x+228422\,{x}^{2}+366806\,{x}^{3} -1591741\,{x}^{4}-8948446\,{x}^{5}
  \nonumber \\
\fl \quad -18137183\,{x}^{6}-10301088\,{x}^{7}+31576074\,{x}^{8}+
82978356\,{x}^{9}+80098415\,{x}^{10} 
 \nonumber \\
\fl \quad -8308172\,{x}^{11}-123518048\,{x}^{12}-158759046\,{x}^{13}
-65285821\,{x}^{14}+78248130\,{x}^{15} 
 \nonumber \\
\fl \quad +152708392\,{x}^{16}+124727752\,{x}^{17}+26488355\,{x}^{18}
-65301174\,{x}^{19}-90679899\,{x}^{20}
  \nonumber \\
\fl \quad -47527872\,{x}^{21}+4032496\,{x}^{22}+27473954\,{x}^{23}
+23107094\,{x}^{24}+9927812\,{x}^{25}
  \nonumber \\
\fl \quad +2288564\,{x}^{26}+245416\,{x}^{27}+10800\,{x}^{28} \Bigr),  
 \nonumber 
\end{eqnarray}
\begin{eqnarray}
\fl p_1 \,=\,\,   
2\, \, (x-1) \cdot 
\Bigl( -1440-15176\,x-3552\,{x}^{2}+632252\,{x}^{3}+3988986\,{x}^{4}+11012538\,{x}^{5} 
 \nonumber \\
\fl \quad +10122851\,{x}^{6}-31358640\,{x}^{7}-125311964\,{x}^{8}
-166380144\,{x}^{9}+20063039\,{x}^{10}
  \nonumber \\
\fl \quad +375202188\,{x}^{11}+523233277\,{x}^{12}+189830162\,{x}^{13}
-422078559\,{x}^{14}-747281488\,{x}^{15}
  \nonumber \\
\fl \quad -440223099\,{x}^{16}+161161298\,{x}^{17}+530901457\,{x}^{18}+491902752
\,{x}^{19}+168466049\,{x}^{20}  \nonumber \\
\fl \quad -168274188\,{x}^{21}-282329480\,{x}^{22}
-158906808\,{x}^{23}-754525\,{x}^{24}+72189798\,{x}^{25} 
 \nonumber \\
\fl \quad +61435092\,{x}^{26}+25677392\,{x}^{27}
+5672988\,{x}^{28}+577984\,{x}^{29}+24000\,{x}^{30} \Bigr),  
 \nonumber 
\end{eqnarray}
\begin{eqnarray}
\fl p_0 \,=\,\,   
-3600-52880\,x-324108\,{x}^{2}-1147996\,{x}^{3}-1575180\,{x}^{4}
+8228874\,{x}^{5}  \nonumber \\
\fl \quad +52977905\,{x}^{6}+108476130\,{x}^{7}-739178\,{x}^{8}
-371064711\,{x}^{9}-563202298\,{x}^{10}  
\nonumber \\
\fl \quad -29824206\,{x}^{11}+842725375\,{x}^{12}+1075242362\,{x}^{13}
+273493047\,{x}^{14}-909934423\,{x}^{15}
  \nonumber \\
\fl \quad -1189246308\,{x}^{16}-414338515\,{x}^{17}+420114304\,{x}^{18}+702981552\,{x}^{19}
+447865799\,{x}^{20}  
\nonumber \\
\fl \quad +30467322\,{x}^{21}-270639170\,{x}^{22}-233990685\,{x}^{23}
-67035676\,{x}^{24}+45089100\,{x}^{25} 
 \nonumber \\
\fl \quad +61580064\,{x}^{26}+29851532\,{x}^{27}
+7030080\,{x}^{28}+714400\,{x}^{29}+28800\,{x}^{30}. 
 \nonumber 
\end{eqnarray}

\section{Analysis of the singular behavior of ${\tilde \chi}^{(3)}_{d}(x)$}
\label{Detailed}

Let us give some detailed analysis of the singular behavior
 of ${\tilde \chi}^{(3)}_{d;2}(x)$ and 
${\tilde \chi}^{(3)}_{d;3}(x)$ around the three singularities:
 $\, x \, \, = \, \, +1, \, -1, \,e^{\pm 2 \pi i/3}$. 

\subsection{Limit of the connection matrix.}
\label{limitcon}

The hypergeometric operator $\, D_z^2 \, +((1+a+b)\, z \, -c) \, D_z \, +a \, b$ has 
two solutions, $u_1\,\,=\,\,\, _2F_1([a,b],\, [c];\, z)$ and
$\,u_2\,= \,\, \,
 z^{1-c} \cdot \,_2F_1([a+1-c,b+1-c],\, [2-c];\, z)$ 
and they connect to $z\,=\,\,1$ (see, for instance, (1) 
on page 108 of~\cite{bateman}) according to the connection
 matrix valid for $\,c\,\neq\, 1$
\begin{eqnarray}
\label{connect}
\hspace{-0.9in}\left[\begin{array}{c}
u_1\\
u_2
\end{array}\right]\,  \,= \,\, \, \, 
C \cdot \left[\begin{array}{c}
_2F_1([a,b],[a+b-c+1];\, 1\,-z)\\
(1-z)^{c-a-b} \cdot \, _2F_1([c-a,c-b],\, [c-a-b+1];\, 1\,-z)
\end{array}\right],
\end{eqnarray}
with
\begin{eqnarray}
C\,\,=\,\,\,
\left[\begin{array}{cc} 
C_{11}&C_{12}\\
C_{21}&C_{22}
\end{array}
\right], \qquad \qquad \quad \hbox{where:}
\end{eqnarray}
\begin{eqnarray}
&&\hspace{-0.5in}C_{11}\,=\,\,\,
\frac{\Gamma(c)\Gamma(c-a-b)}{\Gamma(c-a)\Gamma(c-b)}, 
\qquad \quad \, \,  \, 
\hspace{.1in}C_{12}\,=\,\,\frac{\Gamma(c)\Gamma(a+b-c)}{\Gamma(a)\Gamma(b)}, 
\\
&&\hspace{-0.5in}C_{21}\,=\,\,\,
\frac{\Gamma(2-c)\Gamma(c-a-b)}{\Gamma(1-a)\Gamma(1-b)}, \qquad 
\hspace{.1in}C_{22}\,=\,\,
\frac{\Gamma(2-c)\Gamma(a+b-c)}{\Gamma(a+1-c)\Gamma(b+1-c)}.
\nonumber
\end{eqnarray}

If we now take the limit $\, c\,\rightarrow\, 1$, which is the case of interest
in our problem, the connection
matrix becomes singular. In this limit, solutions with 
$\, \ln(z)$ occur, and we write
$\,u_2\,\,=\,\,\,\, u_1\,\,\, +(1-c)\cdot {\tilde u_2}$,
 where $\,u_1 \,\,=\,\,\, _2F_1([a,b],\, [1];\, z)$ and
\begin{equation}
\hspace{-0.1in}\, \, \, {\tilde u_2}\,\,=\,\,\,
 \ln z \cdot \, _2F_1([a,b],[1]; \, z)\,\, \, 
-\frac{\partial}{\partial c} \, _2F_1([a+1-c,b+1-c],\, [2-c]; \, z)\Big|_{c=1},
\end{equation}
yielding for the connection of the solutions $\,u_1$ and $\, {\tilde u_2}$:
\begin{eqnarray}
\label{con1}
\hspace{-0.8in}\left[\begin{array}{c}
u_1\\
{\tilde u}_2
\end{array}\right]\, \,= \, \,  \, \,
{\tilde C} \cdot \left[\begin{array}{c}
_2F_1([a,b], [a+b]; \, 1\,-z)\\
(1-z)^{1-a-b} \cdot \, _2F_1([1-a,1-b], [2-a-b];\, 1\,-z)
\end{array}\right], 
\end{eqnarray}
with
\begin{eqnarray}
{\tilde C}\,=\,\,\left[\begin{array}{cc} 
C_{11}&C_{12}\\
{\tilde C}_{21}&{\tilde C}_{22}
\end{array}
\right], \qquad \qquad \hbox{where:} \nonumber 
\end{eqnarray}
\begin{eqnarray}
\label{D6}
&&\hspace{-0.8in}{\tilde C}_{21}\,\,\,=\,\,\,\,
\lim_{c\rightarrow 1}\frac{C_{21}-C_{11}}{1-c}
\, \,  =\,\,\, 
\frac{\Gamma(1-a-b)}{\Gamma(1-a)\Gamma(1-b)} \cdot 
\, (2\psi(1)-\psi(1-a)-\psi(1-b)),\nonumber \\
&&\hspace{-0.8in}{\tilde C}_{22}\,\, = \, \,\, \,  \lim_{c\rightarrow  1}
 \, \frac{C_{22}-C_{12}}{1-c}
\,\, =\,\,\, \,  \frac{\Gamma(a+b-1)}{\Gamma(a)\Gamma(b)}
\cdot (2\psi(1)\,-\psi(a)\,-\psi(b)),  
\end{eqnarray}
where $\psi(z)\,=\,\,\,\Gamma'(z)/\Gamma(z)$.

\subsection{The behavior as $x\, \rightarrow\, 1$}

To evaluate $ \,{\tilde \chi}^{(3)}_{d;2}(x)$ 
for $ \,x \,\rightarrow \, 1$ we use the evaluation of 
$\, _2F_1([1/2, \pm1/2],\,[1];\, x^2)$ for $ \,x \,\rightarrow \, 1$, 
and find, from (\ref{chi2}), that its singular part reads
\begin{equation}
\label{chi2x1}
\hspace{-0.5in}{\tilde \chi}^{(3)}_{d;2} (Singular, x=1)
\,\,=\,\,\, \frac{2}{\pi(1-x)^2} \,\, -\frac{1}{\pi(1-x)}
\, + \frac{1}{2 \pi}\, \ln(1-x).
\end{equation}

To evaluate ${\tilde \chi}^{(3)}_{d;3}(x)$ as  $\, x\, \rightarrow\,  1$ we
specialise (\ref{con1}) to $\,a \, = \, 1/6$, $\,b \, = \, 1/3$, 
 $\, z \, = \, \, Q$, where $\, Q$ is defined by (\ref{qdef}). 
 Then as $\,x\,\rightarrow \,1$ one has
$\, (1-Q)^{-1/2} \,\, \rightarrow  \,\,  \,\,\,\, \,
 2/\sqrt 3/(1-x) \,\,\, \, -1/\sqrt 3$, 
and, thus, one deduces the singular part of 
$\,{\tilde \chi}^{(3)}_{d;3}(x)$ using (\ref{Prud}), or
 rather\footnote[2]{Using (\ref{M1}) 
and the fact that $\, _2F_1([1/3,\, 1/6], \, [1]; \, Q)$ is a solution of $\, X_2$.} 
${\tilde \chi}^{(3)}_{d;3}(x) \, = \,$
$\, \, M_1(_2F_1([1/3, \, 1/6], \, [1]; \, Q)^2)$,
 where the linear differential operator $\, M_1$ is defined by (\ref{M1}):
\begin{eqnarray}
\fl \qquad  {\tilde \chi}^{(3)}_{d;3} (Singular, x=1)
\, \,  \, \,   = \, \, \, \, \, \, \frac{6}{\pi} \cdot  \frac{1}{(1-x)^2} \, 
\nonumber \\
\fl \qquad  \qquad  \qquad  
+\frac{3}{(1-x)} \cdot \left[
-\frac{1}{\pi} \, +\frac{5\pi}{9\Gamma^2(5/6)\Gamma^2(2/3)} \, \,
-\frac{8\pi}{\Gamma^2(1/6)\Gamma^2(1/3)}\right]. 
\label{chi3x1}
\end{eqnarray}

\subsection{The behavior as $x\, \rightarrow \, -1$}
\label{minusone}

When $\,x\, \rightarrow\, -1$ it is straightforward from (\ref{chi2}) to
obtain
\begin{equation}
\label{chi2final}
{\tilde \chi}^{(3)}_{d;2}(Singular, x=-1)
\,\,\,=\,\,\,\,\, \frac{1}{2\pi}\cdot \ln (1+x).
\end{equation}

To evaluate $\, {\tilde \chi}^{(3)}_{d;3}$ we note, 
when $\, x\, \rightarrow \, -1$, that $Q$ vanishes as
$Q \,\, \sim \,\, \,\frac{27}{4}\, (1+x)^2$.
However, we cannot directly set $Q\, =\, 0$ in (\ref{chi31}) or (\ref{Prud})
 because we must
analytically connect the solution analytic at $x\, =\, 0$ to the proper
solution at $x\, =\, -1$. To do this we use the results of \ref{limitcon}.
Using the fact that there is no singularity at $x\,=\,-1/2$,
we see that we must choose near $x\,=\,-1/2$
\begin{equation}
(1-Q)^{1/2}\,\,=\,\,\, 
\frac{(1-x) \cdot (1\,+2x) \cdot(2\,+x)}{2\cdot (1+x+x^2)^{3/2}}, 
\end{equation}
which is positive for $\,-1/2\,<\,x\,<\,0$ and negative for $\,-1\,<\,x\,<\,-1/2$.
Therefore, for  $\,-1\,<\,x\,<\,-1/2$, we see that
\begin{eqnarray}
\label{map1}
&&\hspace{-0.7in}u_1\,\, =\,  \,\,
 _2F_1([1/6,1/3],\, [1]; \, Q) \,\,\longrightarrow\,\,
\frac{\Gamma(1/2)}{\Gamma(5/6)\Gamma(2/3)} \cdot \,
_2F_1([1/6,1/3],[1/2];\, 1\,-Q)
\nonumber\\
&&-\frac{\Gamma(-1/2)}{\Gamma(1/6)\Gamma(1/3)}
\cdot \,(1-Q)^{1/2} \cdot \, _2F_1([5/6,2/3],[3/2];\, 1\,-Q).
\end{eqnarray}
Furthermore for $\,-1\,<\,x\,<\,-1/2$
\begin{eqnarray}
&&\hspace{-0.9in}_2F_1([1/6,1/3],\,[1], \, Q) \, \rightarrow  \,\,\,
\frac{{\sqrt 3}}{2\pi}\, {\tilde u}_2\,
+\frac{1}{2}\left(\frac{\Gamma(1/2)}{\Gamma(5/6)\Gamma(2/3)} \cdot \,{\tilde C}_{22}
\,+ \frac{\Gamma(-1/2)}{\Gamma(1/6)\Gamma(1/3)}\,{\tilde C}_{21}\right)\cdot u_1
\nonumber\\
&&\hspace{-0.7in}= \,\, \,\frac{\sqrt{3}}{2\pi}\,
\left(_2F_1([1/6,1/3],\,[1]; \, Q)\,\ln Q \,
-\frac{\partial}{\partial c}\, _2F_1([7/6-c,4/3-c],\,[2-c];\, Q)|_{c=1}\right)
\nonumber\\
&&\hspace{-0.7in}+\frac{1}{2}\left(\frac{\Gamma(1/2)}{
\Gamma(5/6)\Gamma(2/3)} \cdot {\tilde C}_{22}
\,+\frac{\Gamma(-1/2)}{\Gamma(1/6)\Gamma(1/3)} \cdot \, 
{\tilde C}_{21}\right) \cdot \, _2F_1([1/6,1/3],\,[1]; \, Q).
 \end{eqnarray}
By using relations (\ref{D6})  we note that
\begin{eqnarray}
&&\hspace{-0.8in}\frac{\Gamma(1/2)}{\Gamma(5/6)\Gamma(2/3)} \cdot {\tilde C}_{22}
\,\, \,  + \frac{\Gamma(-1/2)}{\Gamma(1/6)\Gamma(1/3)}\cdot  {\tilde C}_{21}
 \, \, \,= \, \,  \, \, \,-\frac{\sqrt 3}{2\pi} \cdot (6\ln 3 +4\ln 2), 
\end{eqnarray}
 and thus as $x \,\rightarrow  \, -1$
\begin{eqnarray}
_2F_1([1/6,1/3],\,[1]; \, Q)\,\,  \rightarrow  \quad  \,\,\, \, \, \, 
\frac{\sqrt 3}{\pi}\, \left(\ln(1+x)\, -2\ln 2\right).
\label{res1}
\end{eqnarray}
Again using  (\ref{Prud}) 
we find, as $\, x\,\rightarrow \,-1$, that
\begin{equation}
\label{chi3final}
{\tilde \chi}^{(3)}_{d;3}\left(Singular, x=-1 \right)\,\,
=\,\,\,\, -\frac{3}{2\pi^2}\cdot \ln^2(1+x)\, \, \,\,
+3\,\frac{2\ln 2-1}{\pi^2} \cdot \, \ln(1+x).
\end{equation}

\subsection{The behavior as $ \,x \, \rightarrow  \,e^{\pm 2 \pi i/3}$}
\label{exppisur3}

When  $x \, \rightarrow \, e^{\pm 2 \pi i/3}$ then 
 $\, Q\, \rightarrow \,  \infty$ and
$\, {\tilde \chi}^{(3)}_{d;3}$ becomes singular. Thus to extract this singularity
we have to connect that solution analytic at $x=0$ to the singularity at
$x\, =\, e^{\pm 2 \pi i/3}$. To do this it is convenient to notice that $Q$
is symmetric about $x\, =\, -1/2$. This is seen by letting 
$\, x\,\, =\,\,\, -1/2\,\, +iy$, to obtain
$\,Q(y)\,\, =\,\,\,\, (1+4y^2)^2/(1-\frac{4}{3}y^2)^3$,  
and defining $z$ by
\begin{equation}
\label{zdef}
z\,=\,\, \, \,(1-Q(y))^{1/2}
\,\, \,=\,\,\, \,\,\frac{i\,y \cdot  \,(9/4+y^2)}{(3/4-y^2)^{3/2}}.
\end{equation}
Furthermore, as
$y$ goes from $0$ to ${\sqrt 3}/2$,   $\, Q(y)$ goes from $1$ to $\infty$.
In the previous section 
we have already connected the solution
analytic at $x\,= \, 0$ with the solution analytic at $x\,=\, -1/2$. 

We rewrite the solutions using (\ref{zdef}) as
\begin{eqnarray}
\label{newsol1}
&&\hspace{-0.8in}_2F_1([1/6,1/3],\,[1]; \, Q) \, \, =\,\,\, \
\frac{\Gamma(1/2)}{\Gamma(5/6)\Gamma(2/3)} \cdot \,
_2F_1([1/6,1/3],\,[1/2]; \, z^2)
\nonumber\\
\label{newsol2}
&&\hspace{-0.3in} +\frac{\Gamma(-1/2)}{\Gamma(1/6)\Gamma(1/3)}
\cdot z \cdot \,
_2F_1([5/6,2/3],\,[3/2]; \, z^2), \\
&&\hspace{-0.8in}F([7/6,4/3],[2];Q)\, =\, \,\, 
18 \cdot \, \frac{\partial}{\partial Q}
   \, _2F_1([1/6,1/3],\,[1]; \, Q).
\nonumber 
\end{eqnarray}
These solutions must  be connected from $y\,=\, 0$ to  
$y\,= \,\, {\sqrt 3}/2$ along the straight line path 
$\, x\,\, =\,\,\, -1/2\,\, +iy$. On
this path $z^2$ is on the negative real axis and, hence, we may use the
connection formula (2) on page 109 of~\cite{bateman}
\begin{eqnarray}
&&\hspace{-0.8in}
_2F_1([a,b],\,[c]; \, z^2)\,\,=\,\,\,
\frac{\Gamma(c)\Gamma(b-a)}{\Gamma(b)\Gamma(c-a)}
\cdot   (-z^2)^{-a} \cdot \, _2F_1([a,1-c+a],[1-b+a]; \, z^{-2})
\nonumber\\
&&\hspace{-0.2in}+\frac{\Gamma(c)\Gamma(a-b)}{
\Gamma(a)\Gamma(c-b)} \cdot \,(-z^2)^{-b}\,\cdot \,
_2F_1([b,1-c+b], [1-a+b]; \, z^{-2}).
\label{con0infty}
\end{eqnarray}
Setting $\, z \,= \,\,\, i \, {\bar z}$
with $\bar z$ real and nonnegative we obtain
\begin{eqnarray}
&&\hspace{-0.8in}_2F_1([1/6,1/3],\,[1];\, Q)
\, =\,\, \,  \frac{\sqrt 3}{2}({\sqrt 3}-i)
\frac{\Gamma(2/3)}{\Gamma(5/6)^2} \cdot 
{\bar z}^{-1/3}\,\cdot \, _2F_1([1/6,2/3 ],\,[5/6]; \, -{\bar z}^{-2})
\nonumber\\
&&\hspace{-0.3in}
+\frac{3}{2}\, (i{\sqrt 3}\, -1) \, \,
 \frac{\Gamma(5/6)^2}{\pi \Gamma(2/3)} \cdot 
{\bar z}^{-2/3} \cdot \, _2F_1([1/3,5/6],\,[7/6]; \, -{\bar z}^{-2}), 
\end{eqnarray}
and a similar relation for $\, _2F_1([7/6,4/3],\,[2]; \, Q)$.
Now we note that $\, 
Q\, \, =\, \, \,1\,-z^2\, \, \,= \, \,  \, 1\, \,+{\bar z}^2$, 
and thus (see (\ref{Prud})) 
\begin{eqnarray}
&&\hspace{-0.8in}_2F_1([1/6,1/3],\,[1]; \, Q)^2 \,\, 
+\frac{2Q}{9} \, _2F_1([1/6,1/3],\, [1]; \, Q) \cdot \, _2F_1([7/6,4/3],\,[2]; \, Q)
\nonumber
\end{eqnarray}
can be rewritten in terms of  $_2F_1$ hypergeometric 
functions of argument $\,-{\bar z}^{-2}$, 
this last expression  going as
$\, -\,54 {\sqrt 3}\cdot  \,i \,\,{\bar z}^{-3}/\pi/35$
when ${\bar z} \, \rightarrow\, \infty$. Noting, as
 $ \,x \,\rightarrow \, e^{2\pi i/3}$, that
\begin{equation}
\Bigl({\bar z},\, \frac{(1\,+2x) \cdot (x\,+2)}{(1-x) \cdot (x^2+x+1)}\Bigr) 
 \,\,\longrightarrow \,\, \,\, \,  
\Bigl(\frac{3^{3/4}}{2} \,e^{-3\pi i/4} \cdot \,
 (x-x_0)^{-3/2}, \,\frac{e^{\pi i/3}} {x-x_0}\Bigr),
\end{equation}
we find that the leading singularity at $x_0 \,= \,\, e^{2\pi i/3}$ 
in $ \, {\tilde \chi}^{(3)}_{d,3}$ is
\begin{eqnarray}
{\tilde \chi}^{(3)}_{d;3}(Singular, x=x_0)  \,  \,&=& \, \, \,  \, \,
\frac{16\cdot 3^{5/4}}{35\pi} \cdot \, e^{\pi i/12} \cdot (x-x_0)^{7/2}.
\end{eqnarray}

\section{Analysis of the singular behavior 
of $\,{\tilde \chi}^{(4)}_{d;2}(t)$ as $t\, \rightarrow\, 1$}
\label{detailchi4}

To get the singular behavior of $\,{\tilde \chi}^{(4)}_{d;2}(t)$
 as $t\, \rightarrow\, 1$, we
use (12) of page 110 of~\cite{bateman}
\begin{eqnarray}
\hspace{-0.9in}_2F_1([1/2,-1/2],\,[1]; \, t) 
\, \rightarrow \, \, \,\, \, \, \frac{2}{\pi} \cdot 
\left(1\, +\frac{1-t}{4} \cdot \, (\ln\frac{16}{1-t} \, -1)\right)
\,\, + \, \cdots 
\end{eqnarray}
and
\begin{eqnarray}
\hspace{-0.9in}_2F_1([1/2,1/2],[1];t)\, \, \,\, 
\longrightarrow  \, \, \,\, \, \, \, 
\frac{1}{\pi} \cdot \, \left(\ln\frac{16}{1-t}\,
\,+\frac{1-t}{4} \cdot \, \ln\frac{16}{1-t}\, -2\right) 
\,\,  \, + \, \cdots 
\end{eqnarray}
Using these in (\ref{chi2}) we find the result quoted 
in the text in (\ref{4chi2final}).

\section{Towards an exact expression for  $\, 3I^>_1\,-4I^>_2$ }
\label{zeta}

The constant $ 3I^>_1\,-4I^>_2$ occurs for ${\tilde \chi}^{(4)}_d(t)$ through
${\tilde \chi}^{(4)}_{d;3}(t)$ which is obtained in (\ref{chi43}) 
by the action of the linear differential
operator $A_3$ on ${}_4F_3([1/2, \,1/2, \,1/2, \,1/2],[1, \,1, \,1]; \, t^2)$.

The constant $ 3I^>_1\,-4I^>_2$ can then be deduced from the $\, 4 \times 4$ 
connection matrix for ${}_4F_3([1/2, \,1/2, \,1/2, \,1/2],[1, \,1, \,1]; \, t^2)$.
The line of the connection matrix relating the solutions
at $\, t \, = \, 0$ to the solutions at $\, t \, = \, 1$ is
\begin{equation}
[A_{4,1},\,\,  \,-1/2\cdot A_{4,1} \, +2/\pi^2, \,\,\, 
A_{4,3} \,-2\, i/\pi, \,\,  \, A_{4,4} \, +i/\pi],
 \nonumber 
\end{equation}
and the constant $ 3I^>_1\,-4I^>_2$ reads
\begin{equation}
\frac{4}{3\pi^2} \cdot (3\, I^>_1\,-4 \, I^>_2) \,  \,\, 
 \,\,\,  = \, \, \,\,\, 
 - \, {{16} \over {\pi^2}} \, + \,{{17} \over {108}} \cdot A_{4,1} \,
  -{{2} \over {3}} \cdot A_{4,3} \,  -\,{{4} \over {3}}\cdot A_{4,4}, 
\end{equation}
The entry $\, A_{4,1}$ of the connection matrix
is actually the {\em evaluation of the hypergeometric function} (\ref{4F3t2}) 
at $\, t \, = \, \, 1$:
\begin{eqnarray}
\label{4F3t2eval}
 A_{4,1} \,\, = \, \,\,  -2 \cdot \,  _4F_3\Bigl(
[ {{1} \over {2}}, \, {{1} \over {2}}, \, {{1} \over {2}}, \,  {{1} \over {2}}], \, 
[1, \, 1, \, 1]; \, 1  \Bigr).  
\end{eqnarray}
There is, at first sight, a ``$\ln(2)$'' coming from
the terms 
in the B\"uhring formula~\cite{buhring} involving the $\psi$
 function. This is the "same"  $\ln(2)$
which appears in the connection formulas for $E(k)$ and $K(k)$.
However, numerically, these $\, \ln(2)$ contributions in
$\, A_{4,3}$ and $\, A_{4,4}$ read respectively 
($\alpha  = \,  -1.9453040783 \, \cdots, \, 
\gamma= \,0.5274495683 \, \cdots$):
\begin{eqnarray}
&&A_{4,3}\, = \, \,  \, \, \alpha \,  \, 
+ \, 2 \cdot \beta \cdot \ln(2), 
\quad \quad \quad 
A_{4,4}\, = \, \,  \, \, \gamma \,  \, - \, \beta \cdot \ln(2),  \\
&& \beta \, \, = \, \,  \, \, 0.101321183  \, \, \cdots 
\nonumber
\end{eqnarray}
The fact that these two entries occur through the
linear combination $\, A_{4,3} \,  +2 \cdot A_{4,4}$ 
actually cancels a  $\, \ln(2)$ contribution in the expression
of the constant $\, 3I^>_1\,-\,4I^>_2$. 

\vskip .2cm 

Similar  constants (see (\ref{I4minus}) for the bulk $\, {\tilde \chi}^{(4)}$,
 (\ref{Clau}) for the bulk $\, {\tilde \chi}^{(3)}$)
can be deduced from entries of the connection matrices
(occurring in the exact calculation of the  
 differential Galois group~\cite{jm4}), such entries being often 
closely related to evaluation, at selected singular points,
 of the holonomic solutions 
we are looking at.
When hypergeometric functions like (\ref{4F3t2}) pop out,
it is not a surprise to have entries that can be simply expressed 
as these  hypergeometric functions at $\, x=\, 1$ (see (\ref{4F3t2eval})).
Along this line, it is worth recalling that $\, \zeta(3)$ 
(or $\, \zeta(5)$, ...) can be simply expressed 
in terms of a simple evaluation at $\, x=\, 1$ of
 a $\, _{q+1}F_q$ hypergeometric 
function~\cite{Krup} (see also~\cite{RivoalThesis}):
\begin{eqnarray}
&&\zeta(3) \,  \, = \, \, \, \, _4F_3([1,1,1,1],[2,2,2]; \, 1),
 \qquad  \\
&&\zeta(5) \,  \, = \, \,  \,\, {{32} \over {31}} \, \cdot \, 
 _6F_5([{{1} \over {2}},{{1} \over {2}},{{1} \over {2}},{{1} \over {2}},{{1} \over {2}},1],
[{{3} \over {2}},{{3} \over {2}},{{3} \over {2}},{{3} \over {2}},{{3} \over {2}}]; \, 1).
\nonumber 
\end{eqnarray}

It is thus quite natural to ask if the sums in
$\, I^>_1$ and $\, I^>_2$ can be evaluated in terms of known constants such as
$\,\zeta(3)$ or evaluations (for instance 
at $\, t\, = \, \, 1$) of hypergeometric functions.

\end{document}